\documentclass[%
aip,
jcp,
amsmath,amssymb,%
reprint,
]{revtex4-2}

\usepackage[pdftex]{graphicx}
\usepackage[version=3]{mhchem}
\usepackage{rotating}
\usepackage{multirow}
\usepackage{physics}

\usepackage{amsmath}
\usepackage{bm,braket}

\newcommand{ \Sec }[1]{Sec.~\ref{sec:#1}}

\newcommand{ \Eq   }[1]{Eq.~(\ref{#1})}
\newcommand{ \Eqs  }[2]{Eqs.~(\ref{#1}) and (\ref{#2})}

\newcommand{ \Table }[1]{Table \ref{tab:#1}}

\newcommand{ \Refer  }[1]{Ref.~\onlinecite{#1}}
\newcommand{ \Refs }[2]{Refs.~\onlinecite{#1} and \onlinecite{#2}}

\newcommand{ \Fig     }[1]{Fig.~\ref{fig:#1}}

\begin{document}

\title{Flexible framework of computing binding free energy using the energy representation theory of solution}

\author{Kazuya Okita}
\affiliation{Division of Chemical Engineering, Graduate School of Engineering Science, Osaka University, Toyonaka, Osaka 560-8531, Japan}
\author{Yusei Maruyama}
\affiliation{Division of Chemical Engineering, Graduate School of Engineering Science, Osaka University, Toyonaka, Osaka 560-8531, Japan}
\author{Kento Kasahara}\email[Author to whom correspondence should be addressed: ]{kasahara@cheng.es.osaka-u.ac.jp}
\affiliation{Division of Chemical Engineering, Graduate School of Engineering Science, Osaka University, Toyonaka, Osaka 560-8531, Japan}
\author{Nobuyuki Matubayasi}\email{nobuyuki@cheng.es.osaka-u.ac.jp}
\affiliation{Division of Chemical Engineering, Graduate School of Engineering Science, Osaka University, Toyonaka, Osaka 560-8531, Japan}

\begin{abstract}
Host-guest binding plays a crucial role in the functionality of various systems, 
and its efficiency is often quantified using the binding free energy, 
which represents the free-energy difference between the bound and dissociated states.
Here, we propose a methodology to compute the binding free energy 
based on the energy representation (ER) theory of solution 
that enables us to evaluate the free-energy difference 
between the systems of interest with the molecular dynamics (MD) simulations.
Unlike the other free-energy methods, such as the Bennett acceptance ratio (BAR), 
the ER theory does not require the MD simulations 
for hypothetical intermediate states connecting the systems of interest, 
leading to reduced computational costs.
By constructing the thermodynamic cycle of the binding process
that is suitable for the ER theory, a robust calculation of the binding free energy is realized.
We apply the present method to the self-association of \textit{N}-methylacetamide (NMA) in different solvents 
and the binding of aspirin to $\beta$-cyclodextrin (CD) in water.
In the former case, the present method estimates 
that the binding free energy decreases as the solvent polarity decreases. 
This trend is consistent with the experimental finding. 
For the latter system, the binding free energies for the two representative CD-aspirin bound complexes, 
primary (P) and secondary (S) complexes, are estimated
to be $-5.2\pm 0.1$ and $-5.03\pm 0.09~\mathrm{kcal~mol^{-1}}$, respectively.
These values are satisfactorily close to those from the BAR method 
[
$-4.2\pm 0.2$ and 
$-4.1\pm 0.2~\mathrm{kcal~mol^{-1}}$ 
for P and S, respectively].
Furthermore, the interaction-energy component analysis reveals 
that the van der Waals interaction between aspirin and CD dominantly contributes to the stabilization of the bound complexes, 
that is in harmony with the well-known binding mechanism in the CD systems.  
\end{abstract}

\maketitle

\section{Introduction\label{sec:intro}}
Host-guest binding has been recognized as one of the most fundamental processes 
in various fields of science. 
For instance, substrate binding to its target protein is a central issue 
in biology because most proteins 
exert their biological functions upon binding.\cite{weinberg2006biology} 
The binding process is also crucial for drug molecules, 
which regulate (promote or inhibit) cellular functions such as cell proliferation mediated by signal transduction.\cite{du2016insights,daviter2021protein}
Molecular dynamics (MD) simulation has played an important role in drug discovery and design\cite{gilson2007calculation, baron2012computational} thanks to its capability of elucidating the binding mechanisms at the atomistic detail based on classical mechanics.
For instance, an inhibitor of HIV integrase was successfully identified through the MD simulations 
combined with molecular docking techniques.\cite{schames2004discovery} 
Binding free energy, the free-energy difference between the bound and dissociated states, 
is regarded as a useful indicator for the efficiency of binding processes and has been extensively 
evaluated through the MD-based approaches.\cite{jorgensen2004many,chodera2011alchemical,De_Vivo_2016, king2021recent}
Therefore, developing methodologies to efficiently compute the binding free energy while enabling systematic analysis 
would be beneficial for in-silico screening of drug candidates.

The thermodynamic integration (TI),\cite{kirkwood1935statistical} free-energy perturbation (FEP),\cite{zwanzig1954high} 
and Bennett acceptance ratio (BAR)\cite{bennett1976efficient} methods
offer a theoretical foundation for estimating the free-energy difference between the two states of interest (endpoints) 
in an exact manner using MD simulations.
In these methods,
the free-energy difference can be evaluated 
by considering a set of intermediate states that connect the endpoint states, which is often referred to as the alchemical pathway.
The double-annihilation scheme (DAS)\cite{jorgensen1988efficient} and double-decoupling scheme (DDS)\cite{gilson1997statistical} are representative approaches for computing the binding free energy using the alchemical pathways.
The DAS describes the binding process 
using the alchemical pathways associated with the gradual vanishing of the guest in both the dissociated state and bound state.
An effective setting of the intermediate states on the pathways is proposed for the DAS.\cite{Fujitani_2005,fujitani2009massively}
The alchemical pathways employed in the DDS are similar to those in the DAS, 
but the restraint is imposed on the guest in the bound state 
to keep it within the binding pocket of the host for all the intermediate states on the pathway.\cite{boresch2003absolute, boresch2024analytical}
Note that the effect of the restraints on the free energy can be removed analytically.
The potential of mean force (PMF) approach coupled with the FEP 
realizes the free-energy calculation 
for large and flexible guest molecules.\cite{woo2005calculation,gumbart2013standard}
The automation 
of the free-energy calculation based on the methodologies mentioned above has expanded the versatility.
\cite{jo2013charmm,heinzelmann2021automation,fu2023standard,fu2021bfee2,kim2020charmm,liu2023accelerating}
However, these methods require conducting the MD simulations for all the intermediate 
states along the alchemical pathway. 
Therefore, the reduction in the computational cost is one of the important subjects. 

The free-energy estimation of binding can be made efficient by adopting approximate methods.\cite{gilson2007calculation,wang2019end,de2020advances} 
The method of linear interaction energy belongs to this category and performs an approximate evaluation of free energy 
from the energetics of the bound and unbound states.\cite{hansson1998ligand} 
The effect of entropy was incorporated by mining local energy minima and quantifying the extents of local fluctuations.\cite{gilson2024rapid} 
The balance between the accuracy and speed of free-energy estimation is pursed when an approximate method is developed.

The classical density functional theory (DFT) of liquids allows for the analytical treatment of alchemical pathways in an approximate manner.\cite{levy2020solvation}
The energy representation (ER) theory of solution is a DFT theory that  
employs the solute-solvent pair interaction energy as a reaction coordinate 
for  effectively describing the relative position and orientation of the solvent molecules around a solute
on one-dimensional space.\cite{matubayasi2000theory,Matubayasi_2019}
This treatment enables us to construct an approximate free-energy functional 
represented in terms of only the information on the endpoint states obtained through the MD simulations, 
leading to the reduction in the computational cost 
as compared to the other alchemical free-energy methods.
The ER theory was formulated to estimate the solvation free energy of a solute, 
which is a free-energy difference associated with the transfer of a solute from the gas phase to the solution phase.
It has proven useful to analyze the solvation energetics for various systems 
including lipid bilayers,\cite{Matubayasi_2019} polymer solutions,\cite{kawakami2012free, kojima2021water} and crystal-surface systems.\cite{tanaka2022crystal} 
Recently, the ER theory has been extended to compute the binding free energy for host-guest systems using an alchemical pathway similar to that of the DDS.
\cite{kasahara2023elucidating}
However, this method is applicable only to host molecules whose holo-form structures resemble 
their apo-form structures.
Since a number of host molecules exhibiting the significant structural changes in the structure due to the binding has been reported,\cite{mobley2007predicting, mobley2007confine, Tang_2017} further development could enhance the applicability and versatility of the ER theory.

Here, we present an ER-based methodology of computing the binding free energy  
applicable to the host-guest systems where the binding event induces a structural change in the host molecule. 
In this approach, the difference in the host structure between the holo- and apo-forms 
is characterized using the distribution on the host-guest interaction energy.
The problematic energy domain in these distributions, which affects the free-energy calculation using the ER theory, is theoretically addressed by introducing a suitable intermediate 
state.
This intermediate state was proposed in a previous study 
on the dissolution of water into polymer membranes.\cite{kawakami2012free}

We apply the developed method to two systems: the self association of \textit{N}-methylacetamide 
(NMA) in different solvents 
and the binding of aspirin to $\beta$-cyclodextrin in water. 
In the first system, 
NMA molecules are known to weakly bind to each other,\cite{schenck2011self} 
allowing for the accurate evaluation of the binding free energy 
through the brute-force MD simulations.
This makes the system suitable for verifying the accuracy of the present method.
In the second system, 
$\beta$-cyclodextrin exhibits different structural populations 
between the apo- and holo-forms.\cite{Tang_2017}, and 
thus it can be used for testing the applicability of the present method.
We also discuss the contributions of the interaction energies between the guest 
and surrounding environments to the binding thermodynamics, aiming to clarify 
the driving force of the binding processes. 
\section{Theory\label{sec:theory}}

\subsection{Theoretical expression of binding free energy}
In this subsection, 
we describe the theoretical expression of the binding free energy 
in terms of solvation free energy.
The reaction scheme for host-guest binding is given by
\begin{align}
\mathrm{H}+\mathrm{G} & \rightleftharpoons\mathrm{B}.
\label{binding_reaction}
\end{align}
Here, H, G, and B signify host, guest and bound complex, respectively.
The equilibrium constant of the above reaction, $K_{\mathrm{a}}$, 
is related with the binding free energy, $\Delta G_{\mathrm{B}}^{\circ}$, as 
\begin{align}
\Delta G^{\circ} & =-\dfrac{1}{\beta}\log c^{\circ}K_{\mathrm{a}}\notag\\
 & =-\dfrac{1}{\beta}\log c^{\circ}\left(\dfrac{\left[\mathrm{B}\right]}{\left[\mathrm{H}\right]\left[\mathrm{G}\right]}\right),
 \label{dGb_def}
\end{align} 
where $c^{\circ}$ is the standard concentration ($c^{\circ} = 1~\mathrm{M}$, typically), and $\left[\mathrm{H}\right]$,
$\left[\mathrm{G}\right]$, and $\left[\mathrm{B}\right]$ 
are the concentrations of H, G, and B, respectively. 
The equilibrium condition of \Eq{binding_reaction} is expressed as     
\begin{align}
\mu_{\mathrm{B}}-\left(\mu_{\mathrm{H}}+\mu_{\mathrm{G}}\right) & =0,
\end{align}
where $\mu_{\mathrm{H}}$, $\mu_{\mathrm{G}}$, and $\mu_{\mathrm{B}}$ are the chemical potentials of H, G, and B, respectively.
The chemical potential of species $\mathrm{S}~(\mathrm{S} = \mathrm{H~or~G})$ is given by\cite{kasahara2023elucidating} 
\begin{align}
\mu_{\mathrm{S}} & =\dfrac{1}{\beta}\log\left(\left[\mathrm{S}\right]\lambda_{\mathrm{S}}\right)\notag\\
 & \quad-\dfrac{1}{\beta}\log\dfrac{{\displaystyle \int d{\bf x}_{\mathrm{S}}\int d{\bf X}_{\mathrm{V}}\,e^{-\beta\left(U_{\mathrm{S}}+U_{\mathrm{SV}}+U_{\mathrm{V}}\right)}}}{V{\displaystyle \int}d{\bf X}_{\mathrm{V}}\,e^{-\beta U_{\mathrm{V}}}},
\label{mu_S}
\end{align}
where $\beta$ is the inverse temperature, $V$ is the system volume,  
and $\lambda_{\mathrm{S}}$ is the kinetic contribution for species S obtained by the integration of the Maxwell-Boltzmann 
velocity distribution.
${\bf x}_{\mathrm{S}}$ is the full-coordinate of species $\mathrm{S}$, and ${\bf X}_{\mathrm{V}}$ 
is the set of the full-coordinates of solvents.
$U_{\mathrm{S}}$, $U_{\mathrm{SV}}$, and $U_{\mathrm{V}}$ 
are the intramolecular energy of $\mathrm{S}$, the interaction energy between S and the solvents, 
and the total potential of the solvents, respectively. 
Regarding species $\mathrm{B}$, the mathematical form of $\mu_{\mathrm{B}}$ is similar to \Eq{mu_S}, but 
the configurational integral over the full-coordinates of H and G, ${\bf x}_{\mathrm{HG}} = \left\{{\bf x}_{\mathrm{H}}, {\bf x}_{\mathrm{G}}\right\}$, needs to be restricted  
to the region corresponding to the bound state.
Let $\Theta_{\mathrm{B}}\left({\bf x}_{\mathrm{HB}}\right)$ be the characteristic function 
whose value is unity when the bound complex is formed and zero otherwise.
Then, $\mu_{\mathrm{B}}$ is expressed as 
\begin{align}
 & \mu_{\mathrm{B}}=\dfrac{1}{\beta}\log\left(\left[\mathrm{B}\right]\lambda_{\mathrm{H}}\lambda_{\mathrm{G}}\right)\notag\\
 & \quad-\dfrac{1}{\beta}\log\dfrac{{\displaystyle \int d{\bf x}_{\mathrm{HG}}\int d{\bf X}_{\mathrm{V}}\,\Theta_{\mathrm{B}}\left({\bf x}_{\mathrm{HG}}\right)e^{-\beta\left(U_{\mathrm{B}}+U_{\mathrm{BV}}+U_{\mathrm{V}}\right)}}}{{\displaystyle V\int d{\bf X}_{\mathrm{V}}\,e^{-\beta U_{\mathrm{V}}}}}, \label{mu_B}
\end{align}
where $d{\bf x}_{\mathrm{HG}} = d{\bf x}_{\mathrm{H}}d{\bf x}_{\mathrm{G}}$.
$U_{\mathrm{B}}$ is the potential of H and G that is composed of the intramolecular energies of H, $U_{\mathrm{H}}$, and 
of G, $U_{\mathrm{G}}$, and the interaction energy between H and G, $U_{\mathrm{HG}}$, as 
\begin{align}
U_{\mathrm{B}} & =U_{\mathrm{H}}+U_{\mathrm{G}}+U_{\mathrm{HG}}.
\end{align}
$U_{\mathrm{BV}}$ is defined as the sum of the interaction energy between H and the solvents, $U_{\mathrm{HV}}$, 
and that between G and the solvents, $U_{\mathrm{GV}}$. 
\begin{align}
U_{\mathrm{BV}} & =U_{\mathrm{HV}}+U_{\mathrm{GV}}.
\end{align}

The solvation free energy of species G represents the change in free energy associated with the solvation process.
This quantity is useful to derive the tractable expression of $\Delta G_{\mathrm{B}}^{\circ}$ from \Eq{dGb_def}, as will be discussed later. 
As for the dissociate state, let us introduce the solution and reference systems whose total potentials are respectively defined as
\begin{align}
\mathcal{V}_{\mathrm{sol}}^{\mathrm{D}} & =U_{\mathrm{G}}+U_{\mathrm{GV}}+U_{\mathrm{V}}, \\
\mathcal{V}_{\mathrm{ref}}^{\mathrm{D}} & =U_{\mathrm{G}}+U_{\mathrm{V}}.
\end{align}
The solvation free energy, $\Delta \mu_{\mathrm{G}}^{\mathrm{D}}$, can be described as 
\begin{align}
\Delta\mu_{\mathrm{G}}^{\mathrm{D}} & =-\dfrac{1}{\beta}\log\dfrac{{\displaystyle \int d{\bf x}_{\mathrm{G}}\int d{\bf X}_{\mathrm{V}}\,e^{-\beta \mathcal{V}_{\mathrm{sol}}^{\mathrm{D}}}}}{{\displaystyle \int d{\bf x}_{\mathrm{G}}\int d{\bf X}_{\mathrm{V}}\,e^{-\beta \mathcal{V}_{\mathrm{ref}}^{\mathrm{D}}}}}.
\label{dmu_D}
\end{align}
Since the interaction between G and the solvents, $U_{\mathrm{GV}}$, is present in $\mathcal{V}_{\mathrm{sol}}^{\mathrm{D}}$ and absent in $\mathcal{V}_{\mathrm{ref}}^{\mathrm{D}}$, 
$\Delta \mu_{\mathrm{G}}^{\mathrm{D}}$ can be interpreted as the free-energy change resulting from the appearance of $U_{\mathrm{GV}}$ for the dissociate state.
Similarly, we define the ``solvation free energy'' of G in the bound complex, $\Delta \mu_{\mathrm{G}}^{\mathrm{B}}$, as 
\begin{align}
\Delta\mu_{\mathrm{G}}^{\mathrm{B}} & =-\dfrac{1}{\beta}\log\dfrac{{\displaystyle \int d{\bf x}_{\mathrm{HG}}\int d{\bf X}_{\mathrm{V}}}\, \Theta_{\mathrm{B}}\left({\bf x}_{\mathrm{HG}}\right)e^{-\beta \mathcal{V}_{\mathrm{sol}}^{\mathrm{B}}}}{{\displaystyle \int d{\bf x}_{\mathrm{HG}}\int d{\bf X}_{\mathrm{V}}}\,\Theta_{\mathrm{B}}\left({\bf x}_{\mathrm{HG}}\right)e^{-\beta \mathcal{V}_{\mathrm{ref}}^{\mathrm{B}}}},
\label{dmu_B}
\end{align}
where $\mathcal{V}_{\mathrm{sol}}^{\mathrm{B}}$ and $\mathcal{V}_{\mathrm{ref}}^{\mathrm{B}}$ are the potentials for the solution and reference systems corresponding to the bound state, respectively, defined as
\begin{align}
\mathcal{V}_{\mathrm{sol}}^{\mathrm{B}} & =U_{\mathrm{G}}+U_{\mathrm{H}}+ U_{\mathrm{HG}} + U_{\mathrm{GV}}+U_{\mathrm{HV}}+U_{\mathrm{V}}, \label{VB_sol} \\
\mathcal{V}_{\mathrm{ref}}^{\mathrm{B}} & =U_{\mathrm{G}}+U_{\mathrm{H}}+U_{\mathrm{HV}}+U_{\mathrm{V}}. \label{VB_ref}
\end{align} 
In \Eq{VB_sol}, all the interactions among G, H, and V are operative, 
and in \Eq{VB_ref}, the interactions between G and H and between G and V are turned off. 
$\Delta \mu_G^{\mathrm{B}}$ is thus the free-energy change 
for introducing the interactions of G with H and V. 
It is called solvation free energy by viewing G as the solute and H and V as the solvent.
The presence of $\Theta_{\mathrm{B}}$ in both the numerator and denominator of \Eq{dmu_B} 
means that the solvation process of species G, which forms the bound complex in both the solution and reference systems, is represented by $\Delta \mu_{\mathrm{G}}^{\mathrm{B}}$. 

Substituting \Eqs{mu_S}{mu_B} into \Eq{dGb_def} yields
\begin{align}
\Delta G^{\circ} 
& =\Delta\mu_{\mathrm{G}}^{\mathrm{B}}-\Delta\mu_{\mathrm{G}}^{\mathrm{D}}+\Delta G_{\mathrm{corr}}^{\circ},
\label{dGb_dmu}
\end{align}
where we have used \Eqs{dmu_D}{dmu_B}, and $\Delta G^{\circ}_{\mathrm{corr}}$ is the standard-state correction term,  
ensuring the concentration of G in the dissociate state is $c^{\circ}$, expressed as 
\begin{align}
&\Delta G_{\mathrm{corr}}^{\circ} \notag \\
&=-\dfrac{1}{\beta}\log \left(c^{\circ}V\dfrac{{\displaystyle \int d{\bf x}_{\mathrm{HG}}\int d{\bf X}_{\mathrm{V}}}\, \Theta_{\mathrm{B}}\left({\bf x}_{\mathrm{HG}}\right)e^{-\beta \mathcal{V}_{\mathrm{ref}}^{\mathrm{B}}}}{{\displaystyle \int d{\bf x}_{\mathrm{HG}}\int d{\bf X}_{\mathrm{V}}}\, e^{-\beta \mathcal{V}_{\mathrm{ref}}^{\mathrm{B}}}}\right). \label{dGcorr}
\end{align}
The configurations of G and those of H and the solvent molecules are independently generated by $U_{\mathrm{G}}$ 
and $U_{\mathrm{H}} + U_{\mathrm{HV}} + U_{\mathrm{V}}$, respectively, in the reference system,  
and thus the logarithm in \Eq{dGcorr} can be computed by the test-particle insertion of G into the configurations of H and the solvent molecules.
Furthermore, $\Delta G_{\mathrm{corr}}^{\circ}$ is intensive and the spatial region for insertion can be made smaller than the simulation cell.\cite{kasahara2023elucidating}

To utilize \Eq{dGb_dmu}, the definition of $\Theta_{\mathrm{B}}$ is needed for $\Delta \mu_{\mathrm{G}}^{\mathrm{B}}$ (\Eq{dGb_dmu}) and $\Delta G_{\mathrm{corr}}^{\circ}$ (\Eq{dGcorr}).
The determination from the shape of the free-energy profile on certain reaction coordinates is a straightforward approach. 
If species G remains inside the binding site of species H during the simulations starting from the bound complex in the solution system, 
$\Theta_{\mathrm{B}}$ can be set to accept all the sampled configurations.
In this case, on the other hand, the explicit form of $\Theta_{\mathrm{B}}$ is needed in the reference system 
to distinguish between the bound complex and others.
The unique determination of $\Theta_{\mathrm{B}}$ is generally impossible except for simple host and guest molecules, 
such as monoatomic molecules, and $\Delta G^{\circ}$ appears to be dependent on the choice of $\Theta_{\mathrm{B}}$ 
through the sampling in the reference system. 
Actually, it can be proved that $\Delta G^{\circ}$ is not affected by the choice of $\Theta_{\mathrm{B}}$ for the reference solvent as described below.
Let us introduce the characteristic function that is different from $\Theta_{\mathrm{B}}$, $\Theta_{\mathrm{B}}^{\prime}$, and the following quantities. 
\begin{align}
\Delta\mu_{\mathrm{G}}^{\mathrm{B}^{\prime}} & =\Delta\mu_{\mathrm{G}}^{\mathrm{B}}-\dfrac{1}{\beta}\log\dfrac{{\displaystyle \int d{\bf x}_{\mathrm{HG}}\int d{\bf X}_{\mathrm{V}}\,\Theta_{\mathrm{B}}\left({\bf x}_{\mathrm{HG}}\right)e^{-\beta \mathcal{V}_{\mathrm{ref}}^{\mathrm{B}}}}}{{\displaystyle \int d{\bf x}_{\mathrm{HG}}\int d{\bf X}_{\mathrm{V}}\,\Theta_{\mathrm{B}}^{\prime}\left({\bf x}_{\mathrm{HG}}\right)e^{-\beta \mathcal{V}_{\mathrm{ref}}^{\mathrm{B}}}}}\notag\\
 & =-\dfrac{1}{\beta}\log\dfrac{{\displaystyle \int d{\bf x}_{\mathrm{HG}}\int d{\bf X}_{\mathrm{V}}}\, \Theta_{\mathrm{B}}\left({\bf x}_{\mathrm{HG}}\right)e^{-\beta \mathcal{V}_{\mathrm{sol}}^{\mathrm{B}}}}{{\displaystyle \int d{\bf x}_{\mathrm{HG}}\int d{\bf X}_{\mathrm{V}}}\, \Theta_{\mathrm{B}}^{\prime}\left({\bf x}_{\mathrm{HG}}\right)e^{-\beta \mathcal{V}_{\mathrm{ref}}^{\mathrm{B}}}},
\label{dmu_B'}
\end{align}
\begin{align}
&\Delta G_{\mathrm{corr}}^{\circ\prime}  =\Delta G_{\mathrm{corr}}^{\circ} \notag \\
& \quad +\dfrac{1}{\beta}\log\dfrac{{\displaystyle \int d{\bf x}_{\mathrm{HG}}\int d{\bf X}_{\mathrm{V}}\,\Theta_{\mathrm{B}}\left({\bf x}_{\mathrm{HG}}\right)e^{-\beta \mathcal{V}_{\mathrm{ref}}^{\mathrm{B}}}}}{{\displaystyle \int d{\bf x}_{\mathrm{HG}}\int d{\bf X}_{\mathrm{V}}\,\Theta_{\mathrm{B}}^{\prime}\left({\bf x}_{\mathrm{HG}}\right)e^{-\beta \mathcal{V}_{\mathrm{ref}}^{\mathrm{B}}}}}\notag\\
 & =-\dfrac{1}{\beta}\log \left(c^{\circ}V\dfrac{{\displaystyle \int d{\bf x}_{\mathrm{HG}}\int d{\bf X}_{\mathrm{V}}}\,\Theta_{\mathrm{B}}^{\prime}\left({\bf x}_{\mathrm{HG}}\right)e^{-\beta \mathcal{V}_{\mathrm{ref}}^{\mathrm{B}}}}{{\displaystyle \int d{\bf x}_{\mathrm{HG}}\int d{\bf X}_{\mathrm{V}}}e^{-\beta \mathcal{V}_{\mathrm{ref}}^{\mathrm{B}}}}\right),
\label{dGcorr'}
\end{align}
By substituting \Eqs{dmu_B'}{dGcorr'} into \Eq{dGb_dmu}, one can rewrite \Eq{dGb_dmu} without any approximations as
\begin{align}
\Delta G_{\mathrm{B}}^{\circ} & =\Delta\mu_{\mathrm{G}}^{\mathrm{B}^{\prime}}-\Delta\mu_{\mathrm{G}}^{\mathrm{D}}+\Delta G_{\mathrm{corr}}^{\circ\prime},
\end{align}
indicating that $\Delta G_{\mathrm{B}}^{\circ}$ does not depend on the choice of $\Theta_{\mathrm{B}}^{\prime}$
when all the terms in the above equation are computed in an exact manner. 
\subsection{Energy representation (ER) theory of solution\label{sec:theory_er}}
The energy representation (ER) theory offers an efficient method for computing solvation free energies using information about the endpoint states. 
In this approach, the full coordinates of the solvents are projected onto the solute-solvent pair interaction energy, 
and the free-energy functional is constructed based on the solvent distribution on the interaction energy, referred to as the energy distribution.
In this subsection, it is our intent here to describe the ER theory only for $\Delta \mu_{\mathrm{G}}^{\mathrm{B}}$ (\Eq{dmu_B}),
as the theoretical developments of the ER theory for $\Delta \mu_{\mathrm{G}}^{\mathrm{D}}$ (\Eq{dmu_D}) have been already reported elsewhere.\cite{Matubayasi_2002, Sakuraba_2014, Matubayasi_2019}

Let $\hat{\rho}_{\alpha}\left(\varepsilon\right)$ denotes the instantaneous distribution for the $\alpha$th species, defined as follows.  
\begin{align}
\hat{\rho}_{\alpha}\left(\varepsilon\right) & =\sum_{i\in\alpha}\delta\left(u_{\alpha}\left({\bf x}_{\mathrm{G}},{\bf x}_{\alpha,i}\right)-\varepsilon\right).
\end{align}
Here, $u_{\alpha}$ is the pair interaction-energy function 
between G and the $\alpha$th species, and ${\bf x}_{\alpha, i}$ is the full-coordinate of
 the $i$th molecule of the $\alpha$th species. $\alpha$ refers to the solvent species (such as water) or H.
By defining the ensemble average in the solution system conditioned by $\Theta_{\mathrm{B}}$ 
and that in the reference system conditioned by $\Theta_{\mathrm{B}}$ respectively as 
\begin{align}
\braket{\cdots}_{\mathrm{sol,}\Theta_{\mathrm{B}}} & =\dfrac{{\displaystyle \int d{\bf x}_{\mathrm{HG}}\int d{\bf X}_{\mathrm{V}}\,\left(\cdots\right)\Theta_{\mathrm{B}}\left({\bf x}_{\mathrm{HG}}\right)e^{-\beta \mathcal{V}_{\mathrm{sol}}^{\mathrm{B}}}}}{{\displaystyle \int d{\bf x}_{\mathrm{HG}}\int d{\bf X}_{\mathrm{V}}\,\Theta_{\mathrm{B}}\left({\bf x}_{\mathrm{HG}}\right)e^{-\beta \mathcal{V}_{\mathrm{sol}}^{\mathrm{B}}}}}, \\
\braket{\cdots}_{\mathrm{ref,}\Theta_{\mathrm{B}}} & =\dfrac{{\displaystyle \int d{\bf x}_{\mathrm{HG}}\int d{\bf X}_{\mathrm{V}}\,\left(\cdots\right)\Theta_{\mathrm{B}}\left({\bf x}_{\mathrm{HG}}\right)e^{-\beta \mathcal{V}_{\mathrm{ref}}^{\mathrm{B}}}}}{{\displaystyle \int d{\bf x}_{\mathrm{HG}}\int d{\bf X}_{\mathrm{V}}\,\Theta_{\mathrm{B}}\left({\bf x}_{\mathrm{HG}}\right)e^{-\beta \mathcal{V}_{\mathrm{ref}}^{\mathrm{B}}}}},
\end{align}
the $\alpha$th solvent distributions in the solution and in the reference system
when the bound complex is formed can be expressed as 
\begin{align}
\rho_{\mathrm{sol,\alpha}}^{\mathrm{B}}\left(\varepsilon\right) & =\braket{\hat{\rho}_{\alpha}\left(\varepsilon\right)}_{\mathrm{sol},\Theta_{\mathrm{B}}},\\
\rho_{\mathrm{ref},\alpha}^{\mathrm{B}}\left(\varepsilon\right) & =\braket{\hat{\rho}_{\alpha}\left(\varepsilon\right)}_{\mathrm{ref},\Theta_{\mathrm{B}}},
\end{align}
respectively.
According to the Kirkwood's charging formula for the alchemical pathway connecting the solution and reference systems through the coupling parameter, $\Delta \mu_{\mathrm{G}}^{\mathrm{B}}$ is expressed using the integral over the coupling parameter.
Introducing the Percus-Yevick (PY)-type and hypernetted-chain (HNC)-type approximations 
against the distributions for the non-endpoint systems 
on the alchemical pathway yields\cite{Matubayasi_2019}
\begin{align}
\Delta\mu_{\mathrm{G}}^{\mathrm{B}} & =\sum_{\alpha}\int d\varepsilon\,\varepsilon\rho_{\mathrm{sol},\alpha}^{\mathrm{B}}\left(\varepsilon\right)\notag\\
 & \quad-\dfrac{1}{\beta}\sum_{\alpha}\int d\varepsilon\,\left(\rho_{\mathrm{sol},\alpha}^{\mathrm{B}}\left(\varepsilon\right)-\rho_{\mathrm{ref},\alpha}^{\mathrm{B}}\left(\varepsilon\right)\right)\notag\\
 & \quad+\dfrac{1}{\beta}\sum_{\alpha}\int d\varepsilon\,\rho_{\mathrm{sol},\alpha}^{\mathrm{B}}\left(\varepsilon\right)\log\dfrac{\rho_{\mathrm{sol},\alpha}^{\mathrm{B}}\left(\varepsilon\right)}{\rho_{\mathrm{ref},\alpha}^{\mathrm{B}}\left(\varepsilon\right)}\notag\\
 & \quad+\mathcal{F}\left[\rho_{\mathrm{sol},\alpha}^{\mathrm{B}}\left(\varepsilon\right),\rho_{\mathrm{ref},\alpha}^{\mathrm{B}}\left(\varepsilon\right),\chi_{\alpha\beta}^{\mathrm{B}}\left(\varepsilon,\eta\right)\right], \label{dmu_functional}
\end{align}
where $\chi_{\alpha\beta}^{\mathrm{B}}\left(\varepsilon, \eta\right)$ is the two-body density-correlation function defined as 
\begin{align}
\chi_{\alpha\beta}^{\mathrm{B}}\left(\varepsilon,\eta\right) & =\braket{\hat{\rho}_{\alpha}\left(\varepsilon\right)\hat{\rho}_{\beta}\left(\eta\right)}_{\mathrm{ref},\Theta_{\mathrm{B}}}\notag\\
 & \quad-\braket{\hat{\rho}_{\alpha}\left(\varepsilon\right)}_{\mathrm{ref},\Theta_{\mathrm{B}}}\braket{\hat{\rho}_{\beta}\left(\eta\right)}_{\mathrm{ref},\Theta_{\mathrm{B}}}.
\end{align}
The first three terms in \Eq{dmu_functional} are the pair free-energy components 
without approximations 
and $\mathcal{F}$ is the approximate free-energy functional for the many-body entropic contributions.
The explicit form of $\mathcal{F}$ is available in \Refer{Matubayasi_2019}.

Evaluating the free energy using \Eq{dmu_functional} is effective 
when the distributions in the solution ($\rho_{\mathrm{sol},\mathrm{\alpha}}^{\mathrm{B}}\left(\varepsilon\right)$) 
and reference ($\rho_{\mathrm{ref},\mathrm{\alpha}}^{\mathrm{B}}\left(\varepsilon\right)$) 
systems overlap well with each other.
However, if the holo-form structures of host molecules observed in the solution system differ from their apo-form structures in the reference system, the distributions \(\rho_{\mathrm{sol},\alpha}^{\mathrm{B}}\left(\varepsilon\right)\) and \(\rho_{\mathrm{ref},\alpha}^{\mathrm{B}}\left(\varepsilon\right)\) may not overlap well.
In such host-guest systems, 
the $\varepsilon$-region with $\rho_{\mathrm{sol,H}}^{\mathrm{B}}\left(\varepsilon\right) \neq 0$ 
and $\rho_{\mathrm{ref},\alpha}\left(\varepsilon\right) = 0$, which is problematic due to 
the integrand of the third term in \Eq{dmu_functional}
\begin{align}
\rho_{\mathrm{sol},\alpha}^{\mathrm{B}}\left(\varepsilon\right)\log\dfrac{\rho_{\mathrm{sol},\alpha}^{\mathrm{B}}\left(\varepsilon\right)}{\rho_{\mathrm{ref},\alpha}^{\mathrm{B}}\left(\varepsilon\right)},
\label{PlogP/P0}
\end{align}
may be too broad, especially in the energy distribution for H ($\rho_{\mathrm{sol,H}}^{\mathrm{B}}\left(\varepsilon\right)$ and $\rho_{\mathrm{ref,H}}^{\mathrm{B}}\left(\varepsilon\right)$).
\begin{figure}[t]
\centering
\vspace*{-\intextsep}
\includegraphics[width=1.0\linewidth]{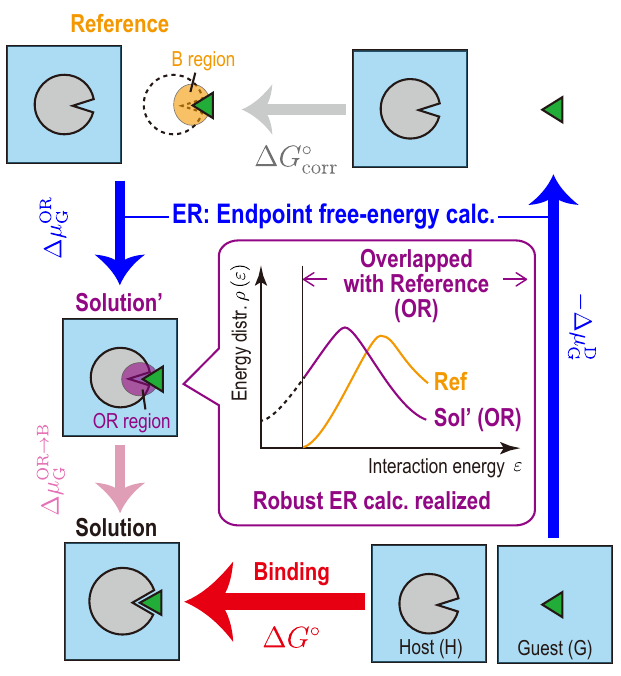}
\caption{Thermodynamic cycle employed in the energy representation (ER) method incorporating a solution state with overlapped distributions with reference (ER-OR). \label{fig:thermo_cycle_ER-OR}}
\end{figure} 
\begin{figure*}[t]
\centering
\vspace*{-\intextsep}
\includegraphics[width=1.0\linewidth]{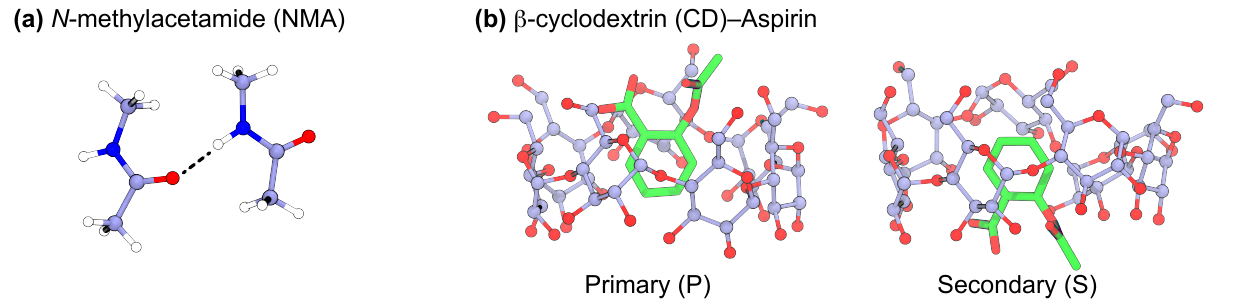}
\caption{Target binding systems. (a) Self-association of \textit{N}-methylacetamide (NMA). (b) $\beta$-cyclodextrin (CD)-aspirin binding. The CD-aspirin complexes in which the hydroxyl group of aspirin points towards the primary and secondary faces are labeled as P and S, respectively.
\label{fig:target_system}}
\end{figure*}
Then, we introduce
a solution state with overlapped distributions with the reference (OR state), 
in which the above problematic $\varepsilon$-region is absent (\Fig{thermo_cycle_ER-OR}).
Such a state can be defined using the following characteristic function. 
\begin{align}
&\Theta_{\mathrm{OR}}\left({\bf x}_{\mathrm{HG}},{\bf X}_{\mathrm{V}}\right) \notag \\ 
&=\Theta_{\mathrm{B}}\left({\bf x}_{\mathrm{HG}}\right)\prod_{\alpha}\prod_{i\in\alpha}\theta\left(\rho_{\mathrm{ref},\alpha}^{\mathrm{B}}\left(u_{\alpha}\left({\bf x}_{\mathrm{H}},{\bf x}_{\alpha,i}\right)\right)\right). \label{Theta_OR}
\end{align}
Here, $\theta\left(x\right)$ is the 
Heaviside's step function given by
\begin{align}
\theta\left(x\right) & =\begin{cases}
0 & x\leq0\\
1 & x>0
\end{cases}.
\end{align}
By defining the free-energy change associated with the transition from B in the reference system
to the OR state in the solution system expressed as 
\begin{align}
\Delta\mu_{\mathrm{G}}^{\mathrm{OR}} & =-\dfrac{1}{\beta}\log\dfrac{{\displaystyle \int d{\bf x}_{\mathrm{HG}}\int d{\bf X}_{\mathrm{V}}}\Theta_{\mathrm{OR}}\left({\bf x}_{\mathrm{HG}},{\bf X}_{\mathrm{V}}\right)e^{-\beta \mathcal{V}_{\mathrm{sol}}^{\mathrm{B}}}}{{\displaystyle \int d{\bf x}_{\mathrm{HG}}\int d{\bf X}_{\mathrm{V}}}\Theta_{\mathrm{B}}\left({\bf x}_{\mathrm{HG}}\right)e^{-\beta \mathcal{V}_{\mathrm{ref}}^{\mathrm{B}}}},
\end{align}
$\Delta \mu_{\mathrm{G}}^{\mathrm{B}}$ can be decomposed into $\Delta \mu_{\mathrm{G}}^{\mathrm{OR}}$ 
and the free-energy change due to the transition from OR to B in the solution system, $\Delta \mu_{\mathrm{G}}^{\mathrm{OR\to B}}$ as
\begin{align}
\Delta\mu_{\mathrm{G}}^{\mathrm{B}} & =\Delta\mu_{\mathrm{G}}^{\mathrm{OR}}+\Delta\mu_{\mathrm{G}}^{\mathrm{OR\to B}},
\end{align}
where $\Delta \mu_{\mathrm{G}}^{\mathrm{OR\to B}}$ is described as
\begin{align}
\Delta\mu_{\mathrm{G}}^{\mathrm{OR}\to\mathrm{B}} & =\dfrac{1}{\beta}\log P_{\mathrm{OR}},
\end{align}
and $P_{\mathrm{OR}}$ is the probability of finding the OR state among the bound-complex configurations, given by
\begin{align}
P_{\mathrm{OR}} & =\dfrac{{\displaystyle \int d{\bf x}_{\mathrm{HG}}\int d{\bf X}_{\mathrm{V}}}\Theta_{\mathrm{OR}}\left({\bf x}_{\mathrm{HG}},{\bf X}_{\mathrm{V}}\right)e^{-\beta \mathcal{V}_{\mathrm{sol}}^{\mathrm{B}}}}{{\displaystyle \int d{\bf x}_{\mathrm{HG}}\int d{\bf X}_{\mathrm{V}}}\Theta_{\mathrm{B}}\left({\bf x}_{\mathrm{HG}}\right)e^{-\beta \mathcal{V}_{\mathrm{sol}}^{\mathrm{B}}}}.
\end{align}
Since $\rho_{\mathrm{sol},\alpha}^{\mathrm{B}}\left(\varepsilon\right)$ 
is the product of $\rho_{\mathrm{ref},\alpha}^{\mathrm{B}}\left(\varepsilon\right)$ 
and a term referring to the solvent-mediated contribution to the potential of mean force, 
in principle $\rho^{\mathrm{B}}_{\mathrm{ref},\alpha}\left(\varepsilon\right) \neq 0$ when $\rho^{\mathrm{B}}_{\mathrm{sol}, \alpha}\left(\varepsilon\right) \neq 0$. 
The difficulty related to \Eq{PlogP/P0} is a practical problem due to finite sampling. 
$\rho^{\mathrm{B}}_{\mathrm{ref},\alpha}\left(\varepsilon\right)$ appearing in \Eq{Theta_OR}
should thus be understood as a numerically computed one, 
and in actual simulations, its argument (energy coordinate $\varepsilon$) 
is discretized to a set of bins with finite widths. 
According to \Eq{Theta_OR}, the OR state is a subset of the B state consisting 
only of the configurations for which all the pair-interaction energies of G 
with each species $\alpha$ fall into energy bins with non-zero $\rho_{\mathrm{ref},\alpha}^{\mathrm{B}}\left(\varepsilon\right)$. If a sampled configuration in the B state contains a pair energy which corresponds to zero $\rho_{\mathrm{ref},\alpha}^{\mathrm{B}}\left(\varepsilon\right)$, that configuration is excluded from OR.
The OR state was called intermediate state in \Refs{kawakami2012free}{Sakuraba_2014}, 
however, to avoid possible confusion with an intermediate state in BAR, 
it is denoted with OR in this work. The interaction of the solute 
with the surroundings is partially turned on in intermediate states of BAR, 
while the solute's interactions are fully turned on in the OR state.
Hereafter, the ER theory incorporating the OR state is referred to as ER-OR.
The ER-OR procedures are schematically depicted in Fig. S1 of the supplementary material. 
\section{Computational methods\label{sec:method}}
\subsection{System setups}
We investigated the self-association of \textit{N}-methylacetamide (NMA) 
in different solvents (acetone, 1,4-dioxane, and chloroform)
and the binding of aspirin to $\beta$-cyclodextrin (CD) in water (\Fig{target_system}).

TIP3P model was used for water, and 
the general Amber force field (GAFF)\cite{wang2004development, wang2006automatic} 
was used for the other species.
Following modeling scheme was adopted to all the species except for water. 
We employed the restrained electrostatic potential (RESP) method\cite{cieplak1995application} 
to determine the point charges 
on the atoms  at HF/6-31G(d) level calculations.
The optimized structures used for the RESP method 
were prepared at MP2/6-31G(d) level calculations except for water and $\beta$-cyclodextrin (CD). 
According to our previous study,\cite{Kasahara_2021} we optimized the CD structure at HF/6-31G(d) level calculations.
The quantum chemical calculations mentioned above were performed with Gaussian 16\cite{frisch2016gaussian} and Antechamber program was used for the RESP method.\cite{case2005amber}
The initial configurations of the systems of interest were built using Packmol.\cite{martinez2009packmol}  

All the simulations were performed with GENESIS 2.0.\cite{jung2015genesis,kobayashi2017genesis,jung2021new}
The Bussi method was used for generating the NVT and NPT ensembles.\cite{Bussi_2007, Bussi_2009}
The velocity Verlet (VVER)\cite{swope1982computer} 
and reversible reference system propagator algorithm (r-RESPA)\cite{tuckerman1992reversible}
integrators were employed for the equilibration and production runs, respectively.
The time intervals for VVER and r-RESPA were $2~\mathrm{fs}$ and $2.5~\mathrm{fs}$, respectively. 
The cutoff distance for the Lennard-Jones (LJ) interactions was $9~\mathrm{\AA}$, 
and smooth particle-mesh Ewald (SPME)\cite{essmann1995smooth} 
was used for computing the electrostatic interactions.
The number of grids for SPME was automatically determined in GENESIS so that 
the grid spacing was shorter than $1.4~\mathrm{\AA}$.
All the bonds that involve hydrogen atoms were constrained with the SHAKE/RATTLE method,\cite{ryckaert1977numerical,andersen1983rattle} 
and water molecules were treated as rigid molecules using SETTLE method.\cite{miyamoto1992settle} 
\subsection{\textit{N}-methylacetamide (NMA) systems}
\subsubsection{Simulation setups\label{sec:NMA_simul_setup}}
For the computation of $\Delta \mu_{\mathrm{G}}^{\mathrm{D}}$ (\Eq{dmu_D}), 
we prepared the trajectories for the system containing an NMA molecule in solvents and for the pure solvent systems,
respectively corresponding to the solution and reference systems for the D state.
For both systems, the box size was $60^{3}~\mathrm{\AA^{3}}$, and the numbers of solvent molecules were set to 1809, 1605, and 1545 
in acetone, 1,4-dioxane, and chloroform, respectively.
The numbers of solvent molecules were determined using the NPT simulations 
to ensure that the system volume fluctuated around $60^{3}~\mathrm{\AA}^{3}$ at 300 K and 1 atm.
For each system, 
we conducted 2 ns NVT simulation for equilibration,  
followed by 10 ns NVT simulation for production.

Regarding $\Delta \mu_{\mathrm{G}}^{\mathrm{B}}$ (\Eq{dmu_B}), 
the trajectories for the solution and reference systems for the B state, 
respectively containing 2 and 1 NMA molecules, are needed. 
As for the latter system (reference),
the trajectory of the solution system for the D state can be used.
In the case of the former systems (solution),
we conducted 2 ns NVT simulation for equilibration,  
followed by 100 ns NVT simulation for production.
These simulations were conducted while applying a following half-flat bottom (HFB) potential 
on the distance between the centers of mass (CoM) of the two NMA molecules ($d$).
\begin{align}
U_{\mathrm{HFB}}\left(d\right) & =\begin{cases}
0 & d<d_{0}\\
k\left(d-d_{0}\right)^{2} & d\geq d_{0}
\end{cases}. \label{HFB} 
\end{align}
Here, $d_{0} = 7~\mathrm{\AA}$ and $k = 10~\mathrm{kcal~mol^{-1}~\AA^{-2}}$.

We performed the NVT simulations for an isolated NMA molecule that is required for the test-particle insertion 
in the ER-based methods.
After 1 ns NVT simulation for equilibration, we performed 1 ns NVT simulation for production. 

To calculate $\Delta G^{\circ}$ using the PMF-based method (exact),\cite{doudou2009standard, Kasahara_2021} 
we also conducted 100 ns NVT simulations in the solution system from the 2 ns equilibration mentioned above 
while applying $U_{\mathrm{HFB}}\left(d\right)$ (\Eq{HFB}) with $d_{0} = 15~\mathrm{\AA}$ and $k=10~\mathrm{kcal~mol^{-1}~\AA^{-2}}$. 
\subsubsection{Binding free-energy calculations\label{sec:nma_setting_bfe}}
The binding free energies, $\Delta G^{\circ}$, were evaluated through the computation
of $\Delta \mu_{\mathrm{G}}^{\mathrm{D}}$, $\Delta \mu_{\mathrm{G}}^{\mathrm{B}}$, and $\Delta G_{\mathrm{corr}}^{\circ}$.
In the case of $\Delta \mu_{\mathrm{G}}^{\mathrm{D}}$,
the energy distributions for the solution and reference systems for the D state 
were computed.
For the reference system, the test-particle insertion was performed 
for computing the distribution, with 1000 insertions for each configuration of the reference system.  
The error estimation of $\Delta \mu_{\mathrm{G}}^{\mathrm{D}}$ 
was done by dividing the solution trajectories into 10 blocks for averaging.

Regarding $\Delta \mu_{\mathrm{G}}^{\mathrm{B}}$, 
the energy distributions in the solution system, 
$\rho_{\mathrm{sol,H}}^{\mathrm{B}}\left(\varepsilon\right)$, was computed 
using the configurations that satisfy the bound-complex criteria.
In this work,
the criteria were defined using the interatomic distances 
involving the oxygen (O) atoms of the carbonyl group and nitrogen (N) 
atoms of the secondary amine.
If the minimum distance among the O-O, N-N, and O-N interatomic distances, 
$d_{\mathrm{min}}$, was shorter than $3.5~\mathrm{\AA}$, 
the NMA dimer was considered a bound complex.
The configurations of the system that satisfy 
this criterion are part of the configurations generated 
with the restraining potential of \Eq{HFB}. $\rho^{\mathrm{B}}_{\mathrm{sol,H}}\left(\varepsilon\right)$
 was constructed by using only those configurations within the distance threshold of $3.5~\mathrm{\AA}$, 
and the other configurations were discarded. See Fig. S2 of the supplementary material for how the choice of the threshold affects the binding free energy.
For the computation of the energy distribution in the reference system, $\rho_{\mathrm{ref,\alpha}}^{\mathrm{B}}\left(\varepsilon\right)$,  
the characteristic function for the B state, $\Theta_{\mathrm{B}}\left({\bf x}_{\mathrm{HB}}\right)$, 
was constructed using the spatial distribution function for the guest NMA, $g\left({\bf r}\right)$, 
and Weeks-Chandler-Andersen (WCA) potential,\cite{weeks1971role} $u_{\mathrm{WCA}}\left({\bf x}_{\mathrm{HB}}\right)$, in addition to $d_{\mathrm{HB}}$.
Here, ${\bf r}$ is the CoM of the guest NMA and $g\left({\bf r}\right)$ was computed using the solution trajectories. 
$\rho_{\mathrm{ref},\alpha}^{\mathrm{B}}\left(\varepsilon\right)$ 
was constructed using the configurations obtained from the test-particle insertion 
that satisfy $d_{\mathrm{min}} \leq 3.5~\mathrm{\AA}$, $g\left({\bf r}\right) > 0$ and $u_{\mathrm{WCA}}\left({\bf x}_{\mathrm{HB}}\right)\leq 15~\mathrm{kcal~mol^{-1}}$.
The same characteristic function was used to perform the test-particle insertion for $\Delta G_{\mathrm{corr}}^{\circ}$.
The number of insertion was 1000 for each configuration of the reference system.
We estimated the statistical error in $\Delta \mu_{\mathrm{G}}^{\mathrm{B}}$ by dividing the solution trajectories into 10 blocks for averaging.  

For comparison, we also computed $\Delta G^{\circ}$ 
using the PMF-based approach.\cite{Kasahara_2021}
In this calculation, only $d_{\mathrm{min}} < 3.5~\mathrm{\AA}$
was used for the bound-complex criteria as well as 
in the calculation of 
$\rho_{\mathrm{sol},\alpha}^{\mathrm{B}}\left(\varepsilon\right)$.
Note that the standard-state concentration was properly treated in this method, 
allowing for a valid comparison of the $\Delta G^{\circ}$ values obtained from this method 
with those from the ER-based methods.  
\subsection{$\beta$-cyclodextrin (CD)-aspirin system}
\subsubsection{Simulation setups}
\begin{table*}[t]
\centering
\vspace*{-\intextsep}
\renewcommand{\arraystretch}{1.4}
\caption{Information on the trajectories used for the free-energy calculations in the CD-aspirin systems. 
The values in parentheses indicate the numbers of replicas for the BAR simulations.
For the BAR simulations, the last 40 ns and 30 ns were used for the D and B states, respectively.\label{tab:traj_info}}
\begin{tabular*}{17cm}{@{\extracolsep{\fill}}cccccccc}
\hline
\hline 
 & \multicolumn{7}{c}{States}\tabularnewline
\cline{2-8} \cline{3-8} \cline{4-8} \cline{5-8} \cline{6-8} \cline{7-8} \cline{8-8} 
 & \multicolumn{3}{c}{D} &  & \multicolumn{3}{c}{B}\tabularnewline
\cline{2-4} \cline{3-4} \cline{4-4} \cline{6-8} \cline{7-8} \cline{8-8} 
 & \# of Traj. & Simul. length  & Total &  & \# of Traj. & Simul. length & Total\tabularnewline
\hline 
BAR & $1\,\left(24\right)$ & $100\,\mathrm{ns}$ & $2400\,\mathrm{ns}$ &  & $10\,\left(57\right)$ & $150\,\mathrm{ns}$ & $85500\,\mathrm{ns}$\tabularnewline
ER, ER-OR (Solution) & $1$ & $20\,\mathrm{ns}$ & $20\,\mathrm{ns}$ &  & $25$ & $20\,\mathrm{ns}$ & $500\,\mathrm{ns}$\tabularnewline
ER, ER-OR (Reference) & $1$ & $10\,\mathrm{ns}$ & $10\,\mathrm{ns}$ &  & $25$ & $20\,\mathrm{ns}$ & $500\,\mathrm{ns}$\tabularnewline
\hline 
\hline 
\end{tabular*}
\end{table*}
We prepared the trajectories required for the computation of $\Delta \mu_{\mathrm{G}}^{\mathrm{D}}$ and $\Delta \mu_{\mathrm{G}}^{\mathrm{B}}$ (\Table{traj_info}).
The simulation scheme was constructed according to our previous study.\cite{Kasahara_2021}
The pure water system composed of 7200 water molecules with the box size of $60^{3}~\mathrm{\AA}^{3}$ 
was built as the reference system for the D state.
After the annealing of the system from 548 K to 298 K during 0.1 ns NVT simulation,
we performed 1 ns NVT simulation for equilibration.
Then, we decided the system size by 1 ns NPT simulation at 1 atm.
The system size at the final step was $60.20^{3}~\mathrm{\AA}^{3}$, 
and this size was used for the other systems described below.  
After further equilibration (0.1 ns NVT), we conducted 10 ns NVT simulation for production.
The solution system for the D state contains an aspirin and 7200 water molecules.
The system was annealed from 548 K to 298 K during 0.1 ns NVT simulation, 
followed by 0.1 ns NVT simulation for equilibration.
Then, we conducted 10 ns NVT simulation for production. 

In the CD-guest systems, 
it is well known that there are two distinct bound complexes, referred to as primary (P) 
and secondary (S) complexes.\cite{khuttan2023taming}
In the P and S complexes, the hydroxyl group of aspirin points towards the primary and secondary faces of CD, respectively. 
We selected 25 different conformations from the trajectories in our previous study for each complex.\cite{Kasahara_2021} 
Using these conformations, 25 initial configurations of the solution system for the B state, 
each containing an aspirin, a CD, and 7200 water molecules were constructed for each complex.
For each initial configuration, we performed 0.1 ns NVT simulation for equilibration 
while imposing the positional restraints on the heavy atoms of the CD and aspirin with the force constant of $1~\mathrm{kcal~mol^{-1}~\AA^{-2}}$. Then, 0.1 ns NVT equilibration was performed.
Following this, we conducted 25 ns NVT production run, and the final 20 ns trajectory was used for analysis.
In the case of the reference system, a CD and 7200 water molecules are involved.
We prepared 25 initial configurations for this system.
Then, we equilibrated the system using 0.1 ns NVT simulation for each configuration, 
followed by 25 ns NVT simulations for production. 
The last 20 ns trajectory was used for analysis.

We also performed the Bennett acceptance ratio (BAR)\cite{bennett1976efficient} simulations for the D and B states 
to compute $\Delta G^{\circ}$ based on the double-annihilation scheme (DAS).
In the case of the D state, 
the initial configuration was taken from the final snapshot after the equilibration for the solution system.
The BAR method combined with Hamiltonian replica-exchange MD (BAR/H-REMD)\cite{jiang2010free} 
implemented in GENESIS\cite{oshima2022modified, matsunaga2022use} 
was performed with a simulation time of 100 ns. 
The setup of the intermediate states (24 states) 
was the same as that in our previous study on the membrane permeation.\cite{yuya2024methodology}
The last 40 ns trajectory for each state was used for the analysis. 
Regarding the B state,
we selected the 10 configurations of the solution system obtained after the 5 ns production simulations for each bound complex.
Then, we conducted the 
150
 ns BAR/H-REMD simulation for each configuration.
The last 30 ns trajectory for each state was used for the analysis.
The intermediate states were defined using the soft-core electrostatic (elec) and van der Waals (vdW) 
interactions with the coupling parameters $\lambda_{\mathrm{elec}}$ (1.000, 0.950, 0.900, 0.850, 0.800, 0.750, 0.700, 0.650, 0.600,0.550, 0.500, 0.450, 0.400, 0.350, 0.300, 0.250, 0.200, 0.150, 0.100, 0.050, and 0.000) 
and $\lambda_{\mathrm{vdW}}$ (1.000, 0.950, 0.900, 0.850, 0.800, 0.750, 0.700, 0.650, 0.600, 
0.550, 0.500, 0.450, 0.400, 0.350, 0.325, 0.300, 0.275, 0.250, 0.225, 0.200, 0.175,
 0.150, 0.140, 0.130, 0.120, 0.110, 0.100, 0.090, 0.080, 0.070, 0.060, 0.050, 0.040, 0.030, 0.020, 0.010, and 0.000).
$\lambda_{i} = 0$ and $1$ correspond to the fully coupled and decoupled states for each interaction-energy component, respectively.
The total number of states is 57.
For all the states, we imposed the HFB potential (\Eq{HFB}) on the distance between the 
CoMs of CD and aspirin, $d$, with $d_{0} = 6~\mathrm{\AA}$ and $k=10~\mathrm{kcal~mol^{-1}~\AA^{-2}}$, 
and on the attractive part of the LJ interaction\cite{Kasahara_2021} between CD and aspirin, $u_{\mathrm{attr}}\left({\bf x}_{\mathrm{HG}}\right)$, defined as
\begin{align}
&U_{\mathrm{FB}}^{\mathrm{uattr}}\left(u_{\mathrm{attr}}\right) \notag \\
&=\begin{cases}
k\left(u_{\mathrm{attr}}-u_{\mathrm{lower}}\right)^{2} & u_{\mathrm{attr}}\leq u_{\mathrm{lower}}\\
0 & u_{\mathrm{lower}}<u_{\mathrm{attr}}\leq u_{\mathrm{upper}}\\
k\left(u_{\mathrm{attr}}-u_{\mathrm{upper}}\right)^{2} & u_{\mathrm{attr}}>u_{\mathrm{upper}}
\end{cases}, 
\end{align}  
The force constant, $k$, was set to $10~\mathrm{kcal^{-1}~mol}$, 
and $\left(u_{\mathrm{lower}},u_{\mathrm{upper}}\right)$ was set to $\left(-38.86,~ -9.35\right)$ for P 
and to $\left(-37.71,~ -11.95\right)$ for S in units of $\mathrm{kcal~mol^{-1}}$.
Note that the values of 
$\left(u_{\mathrm{lower}},u_{\mathrm{upper}}\right)$ were determined
from the lower and upper limits of $u_{\mathrm{attr}}$ observed in the solution systems.
\subsubsection{Binding free-energy calculations}
The scheme for computing $\Delta G^{\circ}$ using the ER-based methods was almost 
parallel to that used for the NMA systems (\Sec{nma_setting_bfe}), 
and the same approach was applied 
for $\Delta \mu_{\mathrm{G}}^{\mathrm{D}}$. 
Therefore, only the settings specific to the computation of $\Delta \mu_{\mathrm{G}}^{\mathrm{B}}$ 
in the CD-aspirin system are described here.
In the simulations for the B state in the solution system, 
we confirmed that aspirin maintained its initial bound pose throughout the simulations. 
Thus, all the configurations generated in the solution system for the B state 
were used to compute $\rho_{\mathrm{ref},\alpha}^{\mathrm{B}}\left(\varepsilon\right)$ for each bound pose.
For the computation of $\rho_{\mathrm{ref},\alpha}^{\mathrm{B}}\left(\varepsilon\right)$ 
and $\Delta G_{\mathrm{corr}}^{\circ}$,
we constructed the characteristic function for the B state, $\Theta\left({\bf x}_{\mathrm{HG}}\right)$, 
in terms of the spatial distribution function for the CoM of aspirin, $g\left({\bf r}\right)$, 
and $u_{\mathrm{attr}}$.
We computed $\rho_{\mathrm{ref},\alpha}^{\mathrm{B}}\left(\varepsilon\right)$ 
using the configurations obtained from the test-particle insertion of aspirin 
into the reference trajectories that satisfy $g\left({\bf r}\right) > 0$ 
and $u_{\mathrm{lower}} \leq u_{\mathrm{attr}}\left({\bf x}_{\mathrm{HG}}\right) \leq u_{\mathrm{upper}}$. 
$\left(u_{\mathrm{lower}},u_{\mathrm{upper}}\right)$ was set to $\left(-38.86,~ -9.35\right)$ for P 
and to $\left(-37.71,~ -11.95\right)$ for S in units of $\mathrm{kcal~mol^{-1}}$.
The number of insertion for each configuration was 10000 for both $\rho_{\mathrm{ref},\alpha}^{\mathrm{B}}\left(\varepsilon\right)$ 
and $\Delta G_{\mathrm{corr}}^{\circ}$.
As noted in the last paragraph of \Sec{theory_er}, 
$\rho_{\mathrm{ref},\alpha}^{\mathrm{B}}\left(\varepsilon\right)$
was constructed by discretizing the energy coordinate $\varepsilon$.
The  bin width of discretization was $0.05~\mathrm{kcal~mol^{-1}}$ in the relevant energy range.
In the case of $\Delta G_{\mathrm{corr}}^{\circ}$, 
the structure of an isolated aspirin was inserted to the spatial region 
containing CD, with the volume of $20^{3}~\mathrm{\AA}$.
The error in $\Delta \mu_{\mathrm{G}}^{\mathrm{B}}$ 
was estimated from the different trajectories for the solution system.  

We also computed $\Delta G^{\circ}$ using the BAR simulations.
Let $\Delta G_{\mathrm{BAR,D}}$ and $\Delta G_{\mathrm{BAR,B}}$ 
represent the free-energy changes along the alchemical pathways in the BAR simulations 
associated with the appearance of the interactions between aspirin and its surrounding environments 
for the D and B states, respectively.
Then, $\Delta G^{\circ}$ can be expressed as
\begin{align}
\Delta G^{\circ} = \Delta G_{\mathrm{BAR,B}} - \Delta G_{\mathrm{BAR,D}} + \Delta G_{\mathrm{BAR,corr}}^{\circ}, 
\end{align}
where $\Delta G_{\mathrm{BAR,corr}}^{\circ}$ is the standard-state correction.
For the computation of $\Delta G_{\mathrm{BAR,B}}$, 
the snapshots satisfying $u_{\mathrm{lower}} \leq u_{\mathrm{attr}} \leq u_{\mathrm{upper}}$, $d \leq 6~\mathrm{\AA}$, 
and the primary/secondary poses criteria for the aspirin's orientation (Fig. S3 of the supplementary material)
were used for the fully coupled ($\lambda_{\mathrm{elec}} = \lambda_{\mathrm{vdW}} = 1$) 
and intermediate states. 
As for the fully decoupled state ($\lambda_{\mathrm{elec}} = \lambda_{\mathrm{vdW}} = 1$), 
the aspirin's orientation was not used for the selecting the snapshots.
According to \Eq{dGcorr}, the standard-state correction, $\Delta G^{\circ}_{\mathrm{BAR, corr}}$, 
is expressed as
\begin{align}
 & \Delta G_{\mathrm{BAR,corr}}^{\circ} \notag \\
 & =-\dfrac{1}{\beta}\log\left(c^{\circ}V\dfrac{{\displaystyle \int d{\bf x}_{\mathrm{G}}\int d{\bf X}_{\mathrm{V}}\,\Theta_{\mathrm{B}}^{\mathrm{BAR}}\left({\bf x}_{\mathrm{HG}}\right)e^{-\beta\mathcal{V}_{\mathrm{ref}}^{\mathrm{B}}}}}{{\displaystyle \int d{\bf x}_{\mathrm{HG}}\int d{\bf X}_{\mathrm{G}}\,e^{-\beta\mathcal{V}_{\mathrm{ref}}^{\mathrm{B}}}}}\right),
\end{align}
where 
\begin{align}
&\Theta_{\mathrm{B}}^{\mathrm{BAR}}\left({\bf x}_{\mathrm{HG}}\right) \notag \\
&=
\begin{cases}
1 &  u_{\mathrm{lower}} \leq u_{\mathrm{attr}} \leq u_{\mathrm{upper}}~\mathrm{and}~ d \leq 6~\mathrm{\AA} \\
0 &  \mathrm{otherwise}
\end{cases}. 
\end{align}
$\Delta G_{\mathrm{BAR,corr}}^{\circ}$ was computed using the test-particle insertion of aspirin 
to the reference trajectories.
The number of insertion was 10000.
The error in $\Delta G_{\mathrm{BAR,D}}$ was estimated by dividing the trajectory of each state 
into 8 blocks for averaging, 
and that in $\Delta G_{\mathrm{BAR,B}}$ was estimated from the different BAR simulation runs.
\section{Results and discussion}
\subsection{Self-association of $N$-methylacetamide (NMA)  in different solvents}
\subsubsection{Energy distribution}
\begin{figure}[t]
\vspace*{-\intextsep}
\includegraphics[width=1.0\linewidth]{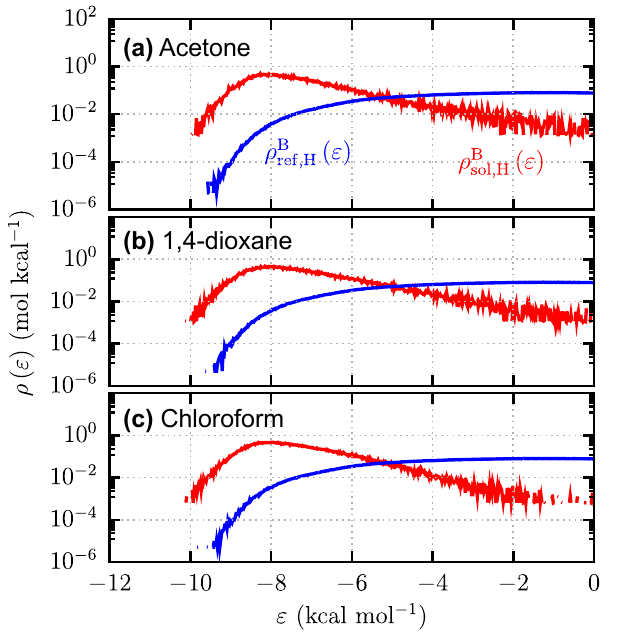}
\caption{Energy distributions of the host NMA molecule in (a) acetone, (b) 1,4-dioxane, and (c) chloroform. \label{fig:nma_engrho}}
\end{figure}
We examine the energy distributions of the host NMA molecule 
in the solution ($\rho_{\mathrm{sol,H}}^{\mathrm{B}}\left(\varepsilon\right)$) 
and reference ($\rho_{\mathrm{ref,H}}^{\mathrm{B}}\left(\varepsilon\right)$) systems 
for the bound complex (\Fig{nma_engrho}).
Note that one of the NMA molecules is regarded as the host, 
while the other is considered guest. 
In acetone (\Fig{nma_engrho}(a)), $\rho_{\mathrm{sol,H}}^{\mathrm{B}}\left(\varepsilon\right)$ 
exhibits a broad peak at $\varepsilon \sim${}$-8.5~\mathrm{kcal~mol^{-1}}$.
As illustrated in \Fig{target_system}(a), 
the bound complex is stabilized by the hydrogen bond between the carbonyl oxygen atom 
and the hydrogen atom in the secondary amine, 
and thus the electrostatic interaction has a dominant contribution to $\rho_{\mathrm{sol, H}}^{\mathrm{B}}\left(\varepsilon\right)$.
We confirm that 
the average value of the interaction energies between the two NMA molecules for the electrostatic component 
is $-6.28\pm 0.01~\mathrm{kcal~mol^{-1}}$, which is significantly larger in magnitude than that for the van der Waals component, 
$-1.049\pm 0.006~\mathrm{kcal~mol^{-1}}$.
It is found that the profile of $\rho_{\mathrm{sol,H}}^{\mathrm{B}}\left(\varepsilon\right)$ 
is largely independent on the solvent species, 
meaning that the distribution of the bound-complex structures is insensitive 
to the surrounding environments.
Similar to $\rho_{\mathrm{sol,H}}^{\mathrm{B}}\left(\varepsilon\right)$, 
$\rho_{\mathrm{ref,H}}^{\mathrm{B}}\left(\varepsilon\right)$ hardly changes its profile 
across the solvent species.
Since the non-overlapping $\varepsilon$-region ($\rho_{\mathrm{sol,H}}^{\mathrm{B}}\left(\varepsilon\right) \neq 0$ 
and $\rho_{\mathrm{ref,H}}^{\mathrm{B}}\left(\varepsilon\right) = 0$) is sufficiently narrow in all the solvent systems,
it is expected that the ``solvation free energy'' in the bound complex, 
$\Delta \mu^{\mathrm{B}}_{\mathrm{G}}$, can be computed using the ER method 
without introducing the OR state (\Eq{Theta_OR}), 
as will be discussed in the next subsection. 
\begin{table}[t]
\centering
\vspace*{-\intextsep}
\renewcommand{\arraystretch}{1.4}
\caption{Binding free energies for the NMA systems obtained through PMF, ER, and ER-OR methods. The errors are provided at the standard error. \label{tab:nma_bfe}}
\begin{tabular}{cccc}
\hline
\hline 
 & \multicolumn{3}{c}{$\Delta G^{\circ}\,\left(\mathrm{kcal\,mol^{-1}}\right)$}\tabularnewline
\cline{2-4} \cline{3-4} \cline{4-4} 
Solvent & PMF & ER & ER-OR\tabularnewline
\hline 
Acetone & $0.8\pm0.1$ & $0.61\pm0.02$ & $0.60\pm0.02$\tabularnewline
1,4-dioxane & $0.1\pm0.1$ & $0.52\pm0.03$ & $0.52\pm0.03$\tabularnewline
Chloroform & $-0.36\pm0.07$ & $0.04\pm0.03$ & $0.03\pm0.04$\tabularnewline
\hline
\hline 
\end{tabular}
\end{table}
\begin{figure}[t]
\vspace*{-\intextsep}
\includegraphics[width=1.0\linewidth]{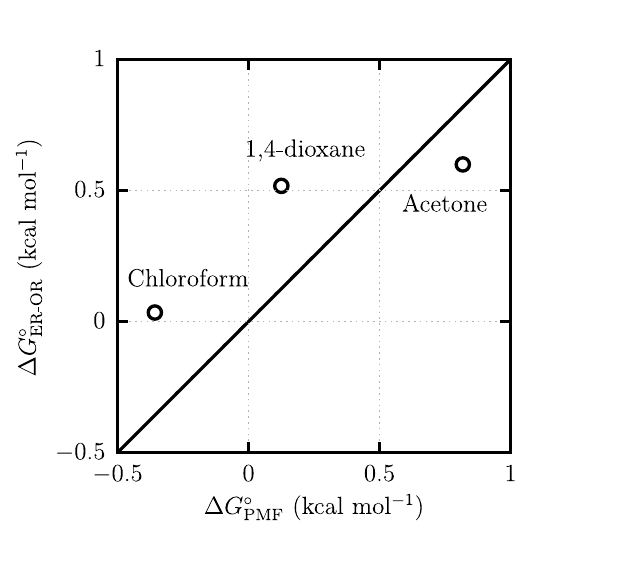}
\caption{Correlation plots of the binding free energies obtained from the ER-OR method, $\Delta G^{\circ}_{\text{ER-OR}}$, against those from the PMF, $\Delta G^{\circ}_{\mathrm{PMF}}$. 
The error bars are not shown in the figure because the standard errors are smaller than $0.1~\mathrm{kcal~mol^{-1}}$ in all the solvent systems. \label{fig:nma_dG_correlation}}
\end{figure}
\subsubsection{Binding free energy}
In this subsection, 
we compare the values of the binding free energies, $\Delta G^{\circ}$, 
obtained from the potentials of mean force (PMF) in an exact way 
with  those from the ER and ER-OR methods to verify the accuracy of the ER-based methods.
The ER and ER-OR methods yield virtually identical $\Delta G^{\circ}$ values in all the solvents examined.
Given that the energy distributions in the solution and reference systems 
($\rho_{\mathrm{sol,H}}^{\mathrm{B}}\left(\varepsilon\right)$ 
and $\rho_{\mathrm{ref,H}}^{\mathrm{B}}\left(\varepsilon\right)$) overlap well, 
it is reasonable for the ER-based methods to yield the identical $\Delta G^{\circ}$ values, regardless of 
the introduction of the overlapped state. 
The values of $\Delta G^{\circ}$ evaluated in this work are listed in \Table{nma_bfe}, 
and the correlation plots of the binding free energies obtained from the ER-OR method, 
$\Delta G^{\circ}_{\mathrm{ER-OR}}$, 
against those from the PMF, $\Delta G^{\circ}_{\mathrm{PMF}}$, are shown in \Fig{nma_dG_correlation}.
The PMF method estimates that $\Delta G^{\circ}$ decreases 
in the order of acetone $>$ 1,4-dioxane $>$ chloroform.
This ordering is consistent with the experimental findings 
that the binding constant for the self-association of NMA increases as solvent polarity decreases.\cite{schenck2011self}
It is seen that the ER and ER-OR methods reproduce the $\Delta G^{\circ}$ ordering predicted from the PMF method.
Furthermore, the deviation from the PMF method is within $0.5~\mathrm{kcal~mol^{-1}}$ in all the solvents. 
Since the NMA dimer has the shallow free-energy minimum in the PMF,
$\Delta G^{\circ}$ is sensitive to the variation in the bound-complex criteria (Fig. S2 of the supplementary material).
However, we confirm that both the ER and PMF methods exhibit the similar behaviors against the variation,
and the $\Delta G^{\circ}$ ordering is not altered.

\subsection{$\beta$-cyclodextrin (CD)-aspirin system}
\subsubsection{Structure and energy distribution of CD}
\begin{figure*}[t]
\centering
\vspace*{-\intextsep}
\includegraphics[width=1.0\linewidth]{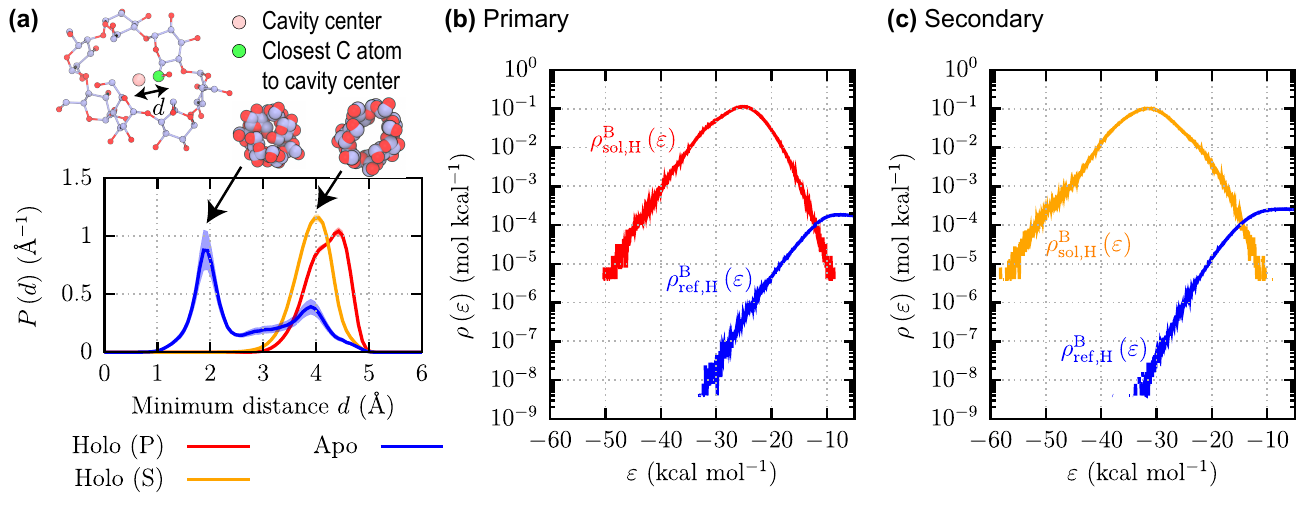}
\caption{Structural difference of CD between apo- and holo-forms, 
and energy distributions of CD. (a) Minimum-distance distribution 
between the cavity center and carbon atoms in CD. 
(b) and (c) Energy distributions of CD in the solution, $\rho_{\mathrm{sol,H}}^{\mathrm{B}}\left(\varepsilon\right)$, and in the reference system, $\rho_{\mathrm{ref,H}}^{\mathrm{B}}\left(\varepsilon\right)$, for complex P (b) and those for complex S (c). 
The cavity center of CD is defined as the center of mass for the ether oxygen atoms.\label{fig:bcd_engrho}}
\end{figure*}
We first assess the impact of binding on the structural population of CD.
\Fig{bcd_engrho}(a) illustrates the distribution of the minimum distance 
between the cavity center of CD and its carbon atoms, denoted as $P\left(d\right)$. 
The cavity center is defined as the center of mass for the ether oxygen atoms in CD.
For the holo-forms of CD in complexes P and S, 
peaks in $P\left(d\right)$ at $d=4.5~\mathrm{\AA}$ and $4~\mathrm{\AA}$, respectively, 
indicate open conformations where the internal cavity of CD is accessible to the guest molecule.
In addition to these open conformations, 
the apo-form exhibits a sharp peak at $d=2~\mathrm{\AA}$, 
reflecting closed conformations where the center of CD is occupied by its own atoms.
In such conformations, one of the sugar rings in CD rotates so that its plane is closer to the cavity center.   
The open and closed conformations in the apo-form are also reported 
by Tang \textit{et al.}\cite{Tang_2017} as well as by Harris \textit{et al.}\cite{Harris:2017aa} 
The localization of $P\left(d\right)$ to the distribution 
corresponding to the open conformations upon binding indicates 
that the structural fluctuations of CD are suppressed by aspirin.

The energy distributions of CD in the solution ($\rho_{\mathrm{sol,H}}^{\mathrm{B}}\left(\varepsilon\right)$) 
and reference systems ($\rho_{\mathrm{ref,H}}^{\mathrm{B}}\left(\varepsilon\right)$) for P and 
those for S are shown in \Fig{bcd_engrho}(b) and (c), respectively.
The peak of $\rho_{\mathrm{sol,H}}^{\mathrm{B}}\left(\varepsilon\right)$ is located at $\varepsilon \sim${}$-25~\mathrm{kcal~mol^{-1}}$ 
for P and $\varepsilon \sim${}$-31.5~\mathrm{kcal~mol^{-1}}$ for S.
Furthermore, the tail of $\rho_{\mathrm{sol,H}}^{\mathrm{B}}\left(\varepsilon\right)$ 
extends further into the negative region for S than for P.
This indicates that the direct interaction between aspirin and CD is stronger for S than for P.
In the case of the reference system,
the difference in $\rho_{\mathrm{ref,H}}^{\mathrm{B}}\left(\varepsilon\right)$ between P and S is 
found to be negligibly small.
The peak position of $\rho_{\mathrm{ref,H}}^{\mathrm{B}}\left(\varepsilon\right)$ 
is $\varepsilon \sim${}$-10~\mathrm{kcal~mol^{-1}}$, and it is shifted in the positive direction 
from that of $\rho_{\mathrm{sol,H}}^{\mathrm{B}}\left(\varepsilon\right)$ ($-25$ and $-31.5~\mathrm{kcal~mol^{-1}}$ for P and S, respectively).
As shown in the profiles of $P\left(d\right)$ (\Fig{bcd_engrho}(a)), 
the accessible $d$-region in the holo-form (solution system) is fully 
covered by that in the apo-form (reference system).
However, a wide non-overlapping region  
between $\rho_{\mathrm{sol,H}}^{\mathrm{B}}\left(\varepsilon\right)$ and 
$\rho_{\mathrm{ref,H}}^{\mathrm{B}}\left(\varepsilon\right)$ is present.
This suggests the presence of a difference 
in the CD structure between the apo- and holo-forms 
that is not captured by $P\left(d\right)$ (\Fig{bcd_engrho}(a)).
\subsubsection{Binding free energy}
\begin{table}[t]
\centering
\vspace*{-\intextsep}
\renewcommand{\arraystretch}{1.4}
\caption{Binding free energies for the CD-aspirin system obtained through BAR, ER, and ER-OR methods. The errors are provided at the standard error.\label{tab:bcd_bfe}}
\begin{tabular}{cccc}
\hline
\hline 
 & \multicolumn{3}{c}{$\Delta G^{\circ}\,\left(\mathrm{kcal\,mol^{-1}}\right)$}\tabularnewline
\cline{2-4} \cline{3-4} \cline{4-4} 
Bound complex & BAR & ER & ER-OR\tabularnewline
\hline 
P & 
$-4.2\pm0.2$ 
& $-3.14\pm 0.09$ & $-5.2\pm 0.1$\tabularnewline
S & 
$-4.1\pm0.2$
& $-2.2\pm0.1$ & $-5.03\pm 0.09$\tabularnewline
\hline
\hline 
\end{tabular}
\vspace*{.2cm} 
\end{table}
\begin{figure}[t]
\centering
\vspace*{-\intextsep}
\includegraphics[width=1.0\linewidth]{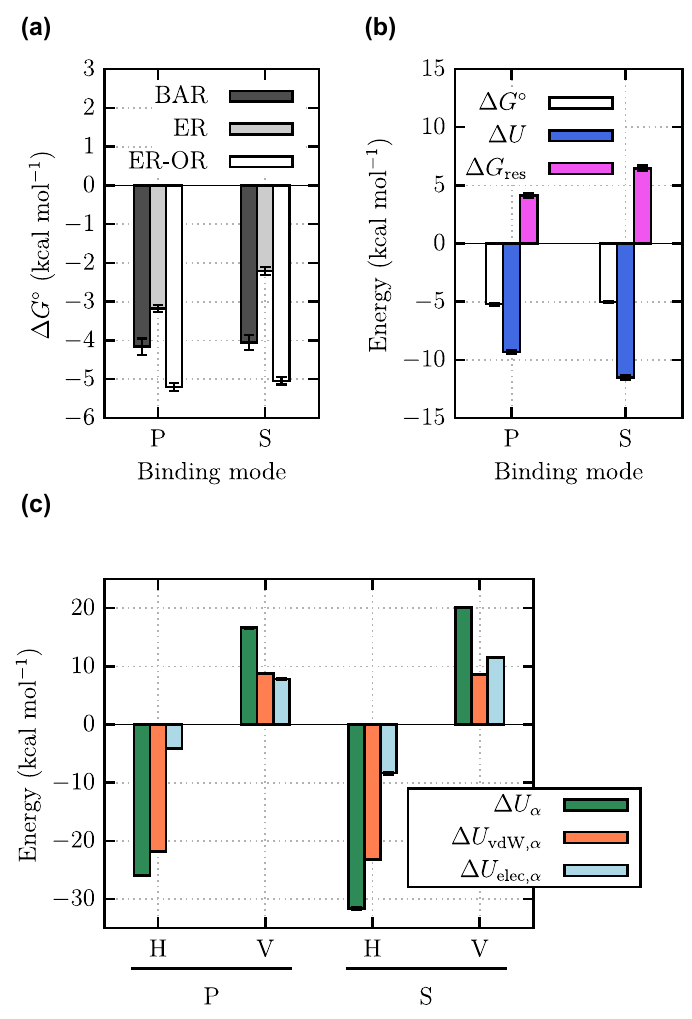}
\caption{Binding free energy, $\Delta G$, and the decomposition analysis based on the interaction energy. 
(a) $\Delta G^{\circ}$ evaluated through the BAR, ER, and ER-OR methods. 
(b) Decomposition of $\Delta G^{\circ}$ into the interaction-energy ($\Delta U$) 
and other ($\Delta G_{\mathrm{res}}$) contributions using \Eq{dG_decomp}. 
In this analysis, the values of $\Delta G^{\circ}$ obtained from the ER-OR methods are used.
(c) Decomposition of $\Delta U$ into the interaction-energy components for each species $\alpha$ ($\alpha = \mathrm{H~(CD)~or~V~(water)}$). The errors are provided at the standard error.\label{fig:bcd_bfe}} 
\end{figure}
We summarize the binding free energies, $\Delta G^{\circ}$, evaluated 
using the BAR, ER, and ER-OR methods in \Fig{bcd_bfe}(a) and \Table{bcd_bfe}.
According to the results from the BAR method, 
the thermodynamic stabilities of complexes P and S are nearly comparable to each other.
In the case of the ER method, the stability of P is predicted 
to be higher than that of S.
The values of $\Delta G^{\circ}$ for S obtained from the ER method differ
by more than $2~\mathrm{kcal~mol^{-1}}$ from those obtained using the BAR method.
%
On the other hand, the ER-OR method reproduces the result revealed by the BAR method 
that the values of $\Delta G^{\circ}$ for P and S are similar to each other.
The improvement achieved by the ER-OR method indicates that the  
introduction of the OR state into the ER method is beneficial 
for robust free-energy calculations, 
when the non-overlapping 
$\varepsilon$-region ($\rho_{\mathrm{sol,H}}^{\mathrm{B}}\left(\varepsilon\right) \neq 0$ 
and $\rho_{\mathrm{ref,H}}^{\mathrm{B}}\left(\varepsilon\right) = 0$) is too broad 
and the interpolation/extrapolation scheme employed in the ER method does not work properly. 
The most time-consuming part of the $\Delta G^{\circ}$ calculation 
for both the ER-OR and BAR methods is the MD simulations for the B state.
The convergence of the computed $\Delta G^{\circ}$ with respect to the simulation timescale is  
much faster with the ER-OR method than with the BAR method (Fig. S4 of the supplementary material).
It is also found that sufficiently long equilibration ($\geq 120~\mathrm{ns}$ in this system) 
is required to obtain the reliable estimates of $\Delta G^{\circ}$ using the BAR method.
According to \Table{traj_info}, the computational cost of these simulations required 
in the ER-OR method is orders-of-magnitude lower than in BAR
given that the error in ER-OR is smaller by a factor of $\sim${}2. 
Introducing a sophisticated scheme of applying the restraint potentials 
in the BAR simulation, such as the virtual bond algorithm (VBA),\cite{boresch2003absolute, boresch2024analytical}  
could accelerate the convergence while maintaining the robustness. 

To clarify the driving force of the binding, 
we elucidate the importance of the interaction energy between aspirin and the surrounding environments
on $\Delta G^{\circ}$.     
According to the endpoint DFT theory, one can decompose the solvation free energy of aspirin, $\Delta \mu_{\mathrm{G}}^{\mathrm{X}}~(\mathrm{X} = \mathrm{B~or~D})$ (\Eqs{dmu_D}{dmu_B}), into the ensemble average of the  
interaction energy between aspirin and its surrounding environments in the solution system at state X, $U^{\mathrm{X}}$, 
and the residual part, $\Delta \mu_{\mathrm{res}}^{\mathrm{X}}$.
Thus, $\Delta G^{\circ}$ can be expressed from \Eq{dGb_dmu} as 
\begin{align}
\Delta G^{\circ} & =\Delta U+\Delta G_{\mathrm{res}},\label{dG_decomp}
\end{align}
where
\begin{align}
\Delta U & =U^{\mathrm{B}}-U^{\mathrm{D}},\\
\Delta G_{\mathrm{res}} & =\Delta\mu_{\mathrm{res}}^{\mathrm{B}}-\Delta\mu_{\mathrm{res}}^{\mathrm{D}}+\Delta G_{\mathrm{corr}}^{\circ}.
\end{align} 
Note that $\Delta G_{\mathrm{res}}$ consists of the contribution from the pair free-energy components, 
the many-body entropic contributions, and the standard-state correction.
\Fig{bcd_bfe}(b) shows the decomposition of $\Delta G^{\circ}$ using \Eq{dG_decomp}.
In this analysis, the values of $\Delta G^{\circ}$ obtained from the ER-OR methods are used.
In both bound complexes, it is found that 
the binding is facilitated by $\Delta U$ and suppressed by $\Delta G_{\mathrm{res}}$,
and that the trend of $\left|\Delta U\right| > \Delta G_{\mathrm{res}}$ 
leads to a negative $\Delta G^{\circ}$.
The value of $\Delta U$ for S is decreased from that for P, 
but this decrease is almost canceled out by the increase of $\Delta G_{\mathrm{res}}$, 
resulting in comparable stability of P and S.
Since the distribution of $P\left(d\right)$ for S is sharper than that for P (\Fig{bcd_bfe}(a)), 
the entropic penalty due to the restriction of the CD structure in S may account 
for the larger value of $\Delta G_{\mathrm{res}}$ for S.

$\Delta U$ is decomposed into the van der Waals and electrostatic interaction-energy components of aspirin with species $\alpha$~($\alpha = \mathrm{H~(CD)~or~V~(water)}$), denoted as $\Delta U_{\mathrm{vdW},\alpha}$ and $\Delta U_{\mathrm{elec},\alpha}$, respectively.
The analysis based on this decomposition is presented in \Fig{bcd_bfe}(c).
Regardless of the complex types, 
the attractive interaction between aspirin and CD, $\Delta U_{\mathrm{H}}$, 
primarily contributes to $\Delta U$ through the van der Waals component, 
$\Delta U_{\mathrm{vdW},\alpha}$. 
This observation is consistent 
with the well-known binding mechanism in which the CD cavity provides a hydrophobic environment, 
enabling guest molecules to be captured through the hydrophobic 
interactions with CD.\cite{schneiderman2000cyclodextrins}
The contribution of the interaction energy between aspirin and water, 
$\Delta U_{\mathrm{V}}$, tends to inhibit binding, 
reflecting the dehydration penalty.
It is observed that both $\Delta U_{\mathrm{vdW,V}}$ and $\Delta U_{\mathrm{elec,V}}$ contribute
almost equally to this penalty.
\section{Conclusion}
In this study, 
we developed a methodology to compute the binding free energies 
based on the energy representation (ER) theory.
The ER theory enables us to calculate the free-energy difference between the two systems of interest, 
referred to as the solution and reference systems.
Unlike other free-energy methods, there is no need 
to conduct the MD simulations for the intermediate states connecting the solution and reference systems, 
leading to the reduction in the computational cost.    
In applications to the calculation of the binding free energy for the host-guest systems ($\Delta G^{\circ}$), however, 
the applicability of the ER theory was limited to the host molecules whose structures 
in the holo-form resemble those in the apo-form. 
In the present method, this problematic structural difference was identified through the distributions on the host-guest interaction 
energy (energy distributions) for the solution and reference systems.
By introducing a solution state involving the overlapped distributions with the reference (OR state), 
we achieved a robust binding free-energy calculation for such host molecules.
The original method is referred to as the ER method, while the present method is referred to as the ER-OR method.
It is noteworthy that, since this state is a subset of the target solution state, 
introducing the additional state into the ER method 
brings no extra computational costs compared to the ER method.

The present method (ER-OR) was first applied to the self-association of \textit{N}-methylacetamide (NMA) 
in different solvents (acetone, 1,4-dioxane, and chloroform).
It was found that $\Delta G^{\circ}$ decreases 
in the order of acetone $>$ 1,4-dioxane $>$ chloroform, 
which aligns with the experimental observations.
Since the energy distribution for the guest NMA 
in the solution and in the reference system overlapped well, 
the $\Delta G^{\circ}$ values evaluated through the ER and ER-OR methods were virtually identical. 
The comparison of the obtained $\Delta G^{\circ}$ values with those 
from the exact method revealed that the differences in $\Delta G^{\circ}$ between the two methods 
were within $0.5~\mathrm{kcal~mol^{-1}}$ in all the solvents.

The binding of aspirin to $\beta$-cyclodextrin (CD) in water was selected as the second target.
In this system, there are two distinct bound complexes, primary (P) and secondary (S) complexes, 
and the CD structure in the holo-form is significantly different from that in the apo-form.
For the bound state, 
the energy distribution of CD for the solution system 
was found to be not overlapped well with those for the reference system 
due to the difference in the CD structures between the holo- and apo-forms.
As a result, the differences in $\Delta G^{\circ}$ between the ER and ER-OR methods
were larger than $1~\mathrm{kcal~mol^{-1}}$ for both P and S.
The ER-OR method reproduced the result revealed by the BAR method that the thermodynamic stabilities 
of P and S are similar to each other,  
indicating an increase in reliability with the introduction of the OR state.
%

The present method works when both of the bound and unbound structures are provided.
Still, it can be employed with any schemes of structure prediction.
For instance, AlphaFold 2/3\cite{jumper2021highly, abramson2024accurate} 
is a choice for preparing a structure which is not known in advance.
The combination of the machine learning (ML)-based schemes of structure prediction
and an all-atom scheme for free-energy evaluation,
such as the present method, will be a promising direction.

Since the computational cost is lower in the present method compared to the other free-energy calculation methods,  
its application to the complex host-guest binding systems appears promising.
For instance, peptide compounds that bind to their target exhibit 
the high flexibility.\cite{muttenthaler2021trends} 
In the present method, the simulations are required only for the endpoint (solution and reference) states, 
allowing for the incorporation of advanced sampling techniques 
to treat such high flexibility, despite their high computational cost.
On the other hand, challenges still remain in the theoretical treatment of the host conformations.
In the present method, 
we assumed the existence of an overlapped region in the energy distributions
between the solution and reference systems. 
However, such a region may be absent in proteins (hosts) 
that exhibit global conformational changes through the induced-fit mechanism.\cite{csermely2010induced}
Introducing an additional state, in which the host molecule has structures 
close to those at the bound state but does not bind the guest, 
into the thermodynamic cycle for $\Delta G^{\circ}$
might be useful for overcoming this challenge.
We believe that the present method and its extensions would be beneficial 
for unveiling binding mechanisms in various host-guest binding systems.  
 
\begin{acknowledgments}
This work is supported by the Grants-in-Aid for Scientific Research (Nos. JP21H05249, JP22J21080, JP23K27313, and JP23K26617)
from the Japan Society for the Promotion of Science, 
the Fugaku Supercomputer Project (Nos. JPMXP1020230325 and JPMXP1020230327), the Data-Driven Material Research Project (No. JPMXP1122714694) from the Ministry of Education, Culture, Sports, Science, and Technology, the Core Research for Evolutional Science and Technology (CREST), Japan Science and Technology Agency (JST) (No. JPMJCR22E3),  
and by Maruho Collaborative Project for Theoretical Pharmaceutics. 
The simulations were conducted using TSUBAME4.0 at Tokyo Institute of Technology, and Fugaku at RIKEN Center for Computational Science through the HPCI System Research Project (Project IDs: hp240223, hp240224, hp240195, and hp240111).
\end{acknowledgments}
\section*{Supplementary material}
The supplementary material contains the protocols for computing binding free energy through 
the ER-based methods, 
the dependency of $\Delta G^{\circ}$ on the definition of the bound state for the NMA systems,
the definition of the primary (P) and secondary (P) poses in the CD-aspirin system,
and the convergence of $\Delta G^{\circ}$ with the trajectory length. 
\section*{Conflict of interest}
The authors have no conflicts to disclose.
\section*{Data Availability}
The source code of ER-OR program and sample data are available at \url{https://github.com/kenkasa/eror-bfe}.
The data that support the findings of this study are available from the corresponding authors upon reasonable request.
\bibliographystyle{apsrev4-1}
\bibliography{bfe}
\clearpage
\widetext

\def\thesection{S\arabic{section}}
\setcounter{section}{0}
\renewcommand{\theequation}{S\arabic{equation}}
\setcounter{equation}{0}
\renewcommand{\thefigure}{S\arabic{figure}}
\setcounter{figure}{0}
\renewcommand{\thetable}{S\arabic{table}}
\setcounter{table}{0}
\renewcommand{\thepage}{S\arabic{page}}
\setcounter{page}{0}

\begin{center}
\Large \bf
Supplement for ``Flexible framework of computing binding free energy using the energy
representation theory of solution''
\end{center}
\section{supplementary figures}
%
%
\begin{figure}[h]
\centering
\includegraphics[width=0.70\linewidth]{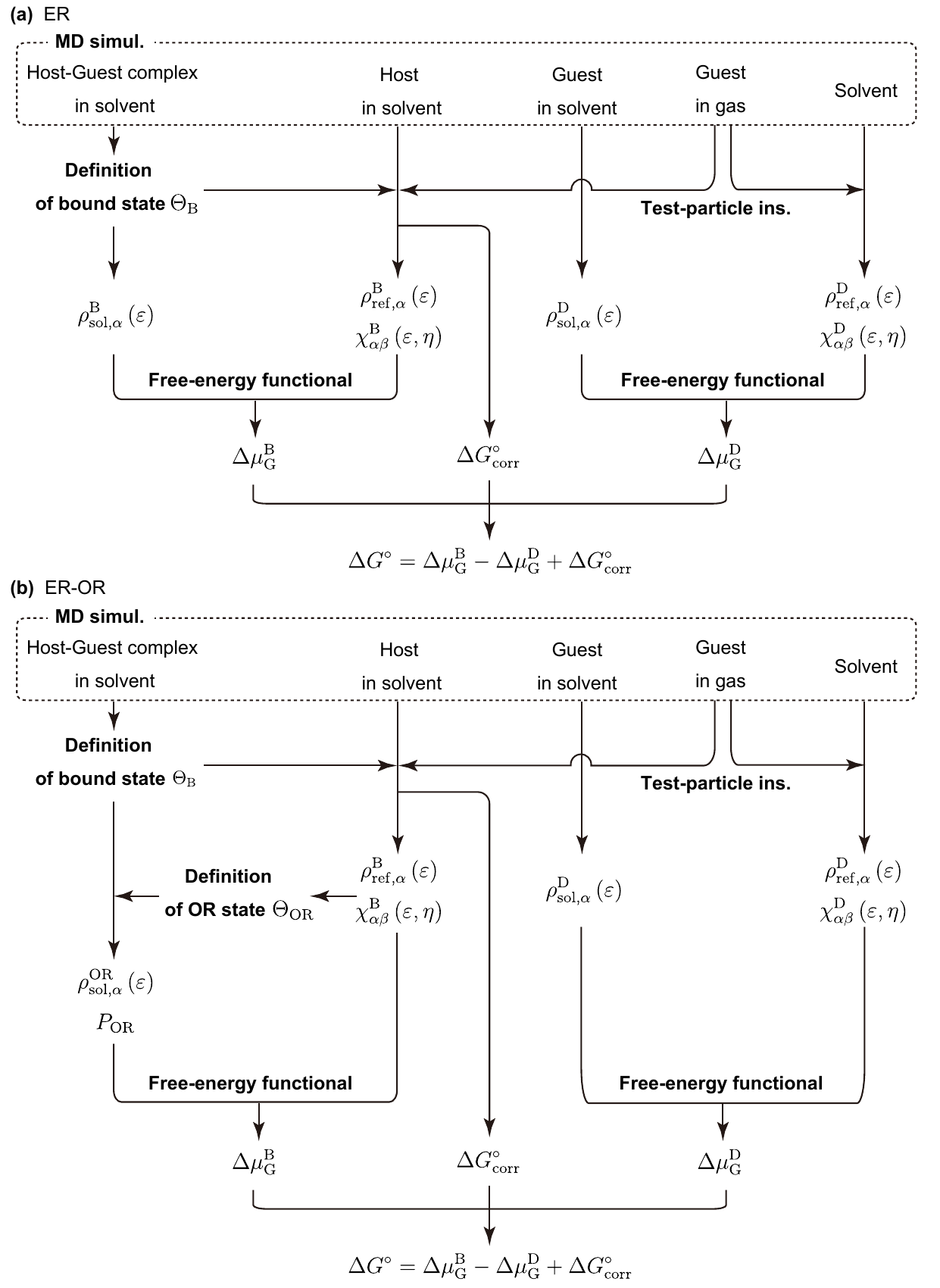}
\caption{Protocols for computing binding free energy, $\Delta G^{\circ}$, through (a) energy-representation (ER) and (b) 
ER incorporating a solution state with overlapped distributions with reference (ER-OR). 
$\rho^{\mathrm{B}}_{\mathrm{sol},\alpha}\left(\varepsilon\right)$, $\rho^{\mathrm{D}}_{\mathrm{sol},\alpha}\left(\varepsilon\right)$, and $\rho^{\mathrm{OR}}_{\mathrm{sol},\alpha}\left(\varepsilon\right)$ are the energy distributions for the bound (B), dissociate (D), and OR states, respectively. $\chi_{\alpha\beta}^{\mathrm{X}}\left(\varepsilon, \eta\right)$ ($\mathrm{X} = \mathrm{B~or~D}$) is the two-body density-correlation function, $P_{\mathrm{OR}}$ is the probability of finding the OR state in the B state, $\Delta \mu_{\mathrm{G}}^{\mathrm{X}}$ ($\mathrm{X} = \mathrm{B~or~D}$) is the solvation free energy of guest in the X state, and $\Delta G^{\mathrm{\circ}}_{\mathrm{corr}}$ is the standard-state correction.\label{fig:eror-scheme}} 
\end{figure}
\begin{figure}[h] 
\centering
\includegraphics[width=1.0\linewidth]{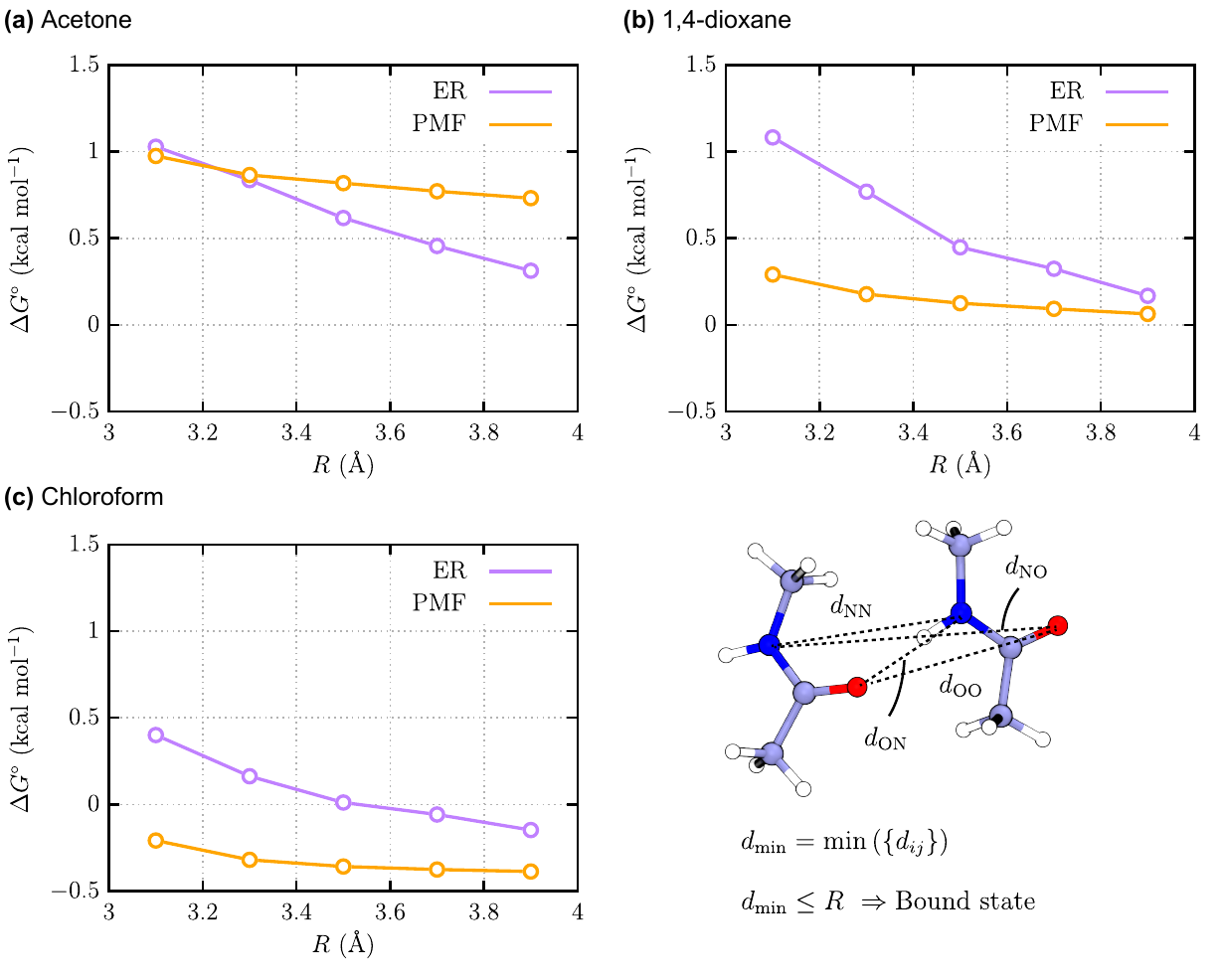}
\caption{
Dependency of $\Delta G^{\circ}$ on the definition of the bound state for the NMA systems.
(a) acetone, (b) 1,4-dioxane, and (c) chloroform.}
\end{figure}
\newpage
\begin{figure}[h] 
\centering
\includegraphics[width=1.0\linewidth]{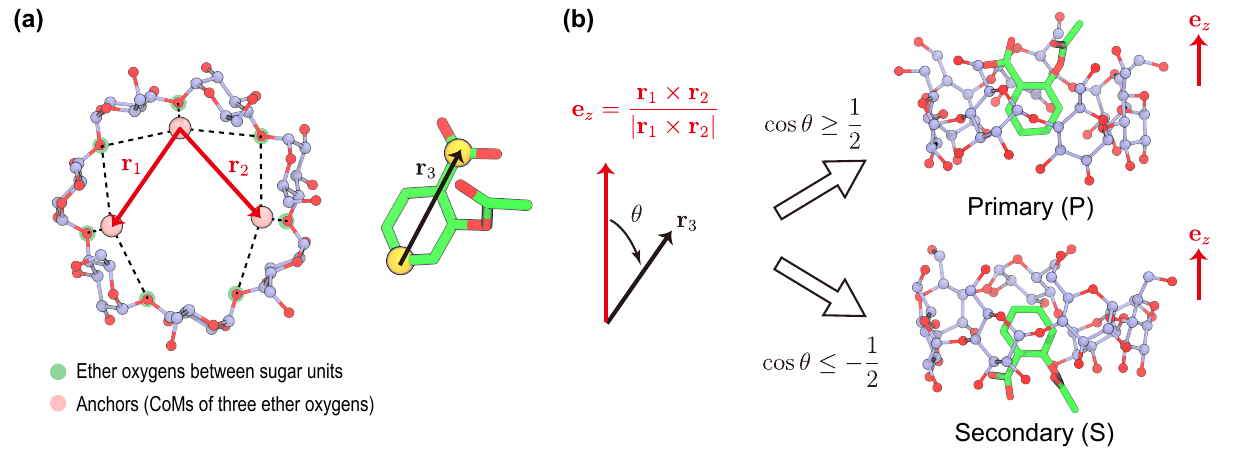}
\caption{Definition of the primary (P) and secondary (S) poses in the $\beta$-cyclodextrin (CD)-aspirin system. 
(a) The vectors on the cross section of CD, ${\bf r}_{1}$ and ${\bf r}_{2}$, and the molecular axis of aspirin, ${\bf r}_{3}$.
${\bf r}_{1}$ and ${\bf r}_{2}$ are the vectors connecting the three anchor particles defined based on the centers of mass (CoMs) of the ether oxygens between the sugar units. The ether oxygens used for defining each anchor particle are indicated as the dash lines in the figure. ${\bf r}_{3}$ is defined using the two carbon atoms in aspirin illustrated as the yellow particles. 
(b) $z$-direction (${\bf e}_{z}$) is defined as the vector product of ${\bf r}_{1}$ and ${\bf r}_{2}$. The angle between ${\bf e}_{z}$ and ${\bf r}_{3}$ is defined as $\theta$, and the P and S poses are defined as the complexes that satisfy $\cos \theta \geq 1/2$ and $\cos \theta \leq - 1/2$, respectively. 
}
\end{figure}
\newpage
\begin{figure}[h]
\centering
\includegraphics[width=1.0\linewidth]{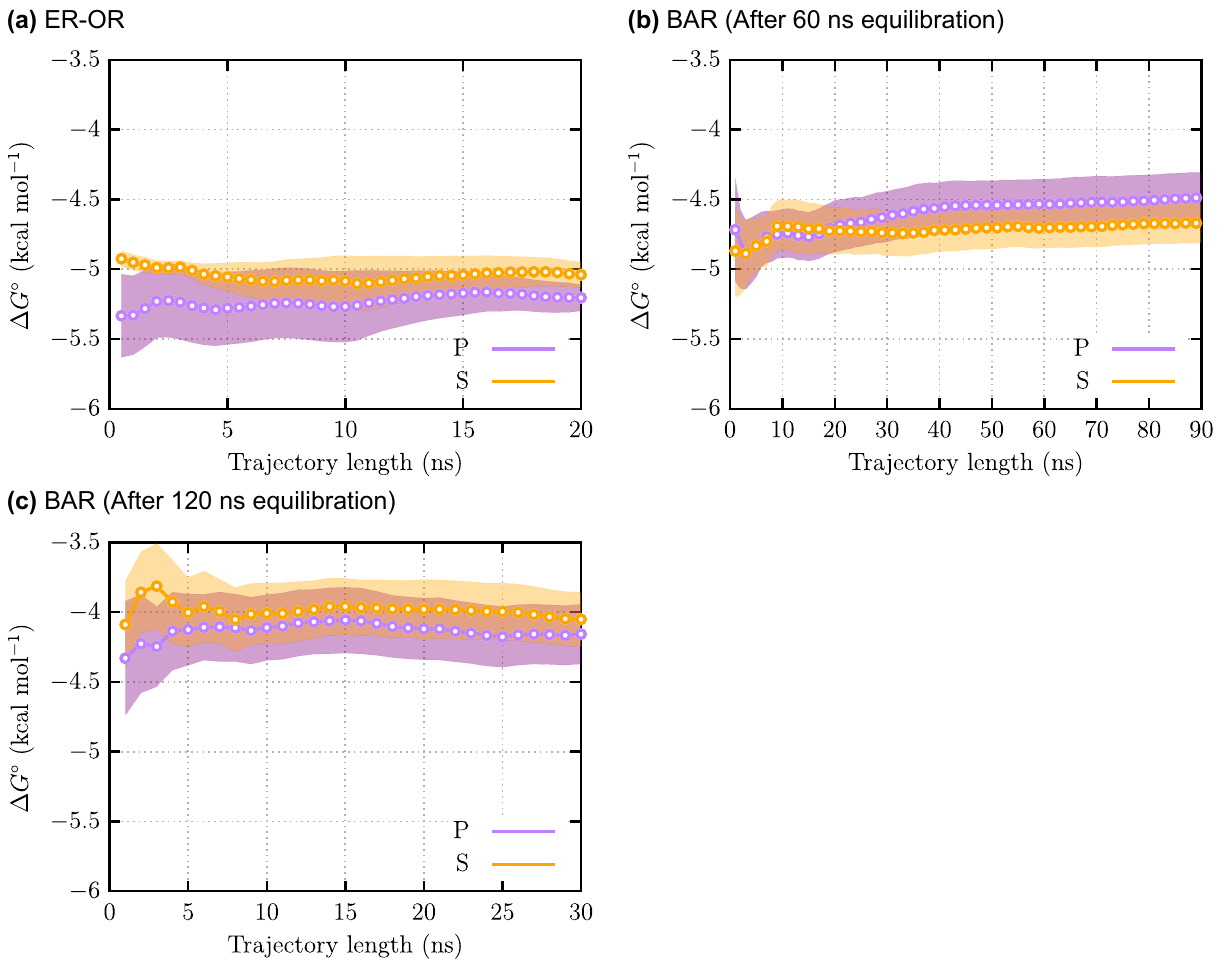}
\caption{Convergence of $\Delta G^{\circ}$ with the trajectory length.
(a) ER-OR, (b) BAR (after 60 ns equilibration), and (c) BAR (after 120 ns equilibration).} 
\end{figure}

\end{document}